\documentclass[fake,dvipsnames]{imsart}

\RequirePackage[OT1]{fontenc}
\usepackage{amsthm,amsmath,amssymb}

\usepackage{amsfonts,graphicx,multirow,bm,tabularx,tikz,lineno}
\usepackage{algorithm}
\usepackage[noend]{algpseudocode}
\usepackage[margin=1.25in,bottom=1.5in]{geometry}

\theoremstyle{definition}
\newtheorem{definition}{Definition}[section]

\usepackage{natbib}
\startlocaldefs
\numberwithin{equation}{section}
\theoremstyle{plain}

\endlocaldefs

\begin{document}

\begin{frontmatter}
\title{Prediction scoring of data-driven discoveries for reproducible research
} 
\runtitle{Prediction Scoring}

\begin{aug}
\author{\fnms{Anna L.} \snm{Smith}$^1$},
\author{\fnms{Tian} \snm{Zheng}$^2$},
\and
\author{\fnms{Andrew} \snm{Gelman}$^2$}

\runauthor{A. Smith, T. Zheng and A. Gelman}

\affiliation{\small $^1$University of Kentucky, $^2$Columbia University}

\end{aug}

\begin{abstract}

Predictive modeling uncovers knowledge and insights regarding a hypothesized data generating mechanism (DGM). Results from different studies on a complex DGM, derived from different data sets, and using complicated models and algorithms, are hard to quantitatively compare due to random noise and statistical uncertainty in model results. This has been one of the main contributors to the {\em replication crisis} in the behavioral sciences.

The contribution of this paper is to apply prediction scoring to the problem of comparing two studies, such as can arise when evaluating replications or competing evidence.

We examine the role of predictive models in quantitatively assessing agreement between two datasets that are assumed to come from two distinct DGMs.
We formalize a distance between the DGMs that is estimated using cross validation.  We argue that the resulting prediction scores depend on the predictive models created by cross validation.  In this sense, the prediction scores measure the distance between DGMs, along the dimension of the particular predictive model.
Using human behavior data from experimental economics, we demonstrate that prediction scores can be used to evaluate preregistered hypotheses and provide insights comparing data from different populations and settings.
We examine the asymptotic behavior of the prediction scores using simulated experimental data and demonstrate that leveraging competing predictive models can reveal important differences between underlying DGMs.
Our proposed cross-validated prediction scores are capable of quantifying differences between unobserved data generating mechanisms and allow for the validation and assessment of results from complex models. 
\end{abstract}


\end{frontmatter}



\section{Introduction} \label{sec.Intro}

Many important scientific advances begin with exploratory investigations of observed data, but much of scientific practice relies on confirmatory analyses that evaluate data against a scientific hypothesis, accounting for uncertainty \citep{tukey_1972}.
In recent years, the scientific community has become increasingly more open to statisticians’ increasingly vocal warnings about this heavy reliance on null hypothesis statistical tests (NHST), and their $p$-values, that constitute much of confirmatory analyses \citep[e.g.,][]{ziliak_mccloskey_2008,nuzzo_2014, wasserstein_lazar_2016, jeske_2019}.
As a result, we have witnessed increasing research interest in alternatives to $p$-values and, more broadly, in methodology that can comprehensively account for the nuances of complex data, complex models, and the increasingly complicated algorithms necessary to estimate these models.
Motivated by this discussion, we consider the statistical problem of exploring and understanding the behavior of a data-driven discovery across \textit{two} sets of observed data that are believed to have been generated from similar processes or models \citep[e.g., a study and its replication, as in][or a pilot study and realized experimental data]{pawel_held_2022}.

A key motivating example is the evaluation of preregistered hypotheses \citep{humphreys_sanchez_vanderwindt_2013, gelman_2013}, often called {\em prediction scoring}, a research planning strategy which arose out of the frustration with $p$-values.
Preregistration requires that researchers make publicly recorded predictions, often based on prior or pilot data, for the scientific hypotheses that will be assessed once the data has been collected.
This forces researchers to clearly differentiate between confirmatory and exploratory analyses, which ensures that $p$-values for the confirmatory analyses can be safely interpreted as intended.
After data collection, the researchers are faced with a natural question:  {\em How well do the preregistered predictions align with the observed data?}
This question requires a method of {\em scoring} the predictions, in the face of the materialized observations and the noise that comes with them.

This process of preregistration presents a unique statistical challenge:  when we preregister our hypotheses we are making assumptions about the underlying data generating mechanism (DGM; e.g., it behaves like it did in this previous study; or we suspect this or that parameter to have a positive or negative significant effect).
Commonly, preregistered studies formulate hypotheses in the form of NHSTs (e.g., about regression coefficients) and predict whether or not the associated $p$-values are significant; the preregistered predictions are then evaluated on a purely binary scale:  is the $p$-value significant or not?
Our proposed prediction scoring approach \textit{quantifies} the differences between preregistered predictions (which are often informed by pilot data) and the realized experimental data; and results in a natural way to visualize these differences.
Our approach represents a unique statistical perspective on preregistration that has gone largely unaddressed.
Furthermore, prediction scores measure differences between DGMs, such as in the preregistration setting but also in more general cases.
For example, prediction scores can leverage competing models to make important discoveries about the underlying DGMs, as in the simulated data examples investigated in Section \ref{sec.SimStudy}.


\section{Our approach} \label{sec.OurApproach}

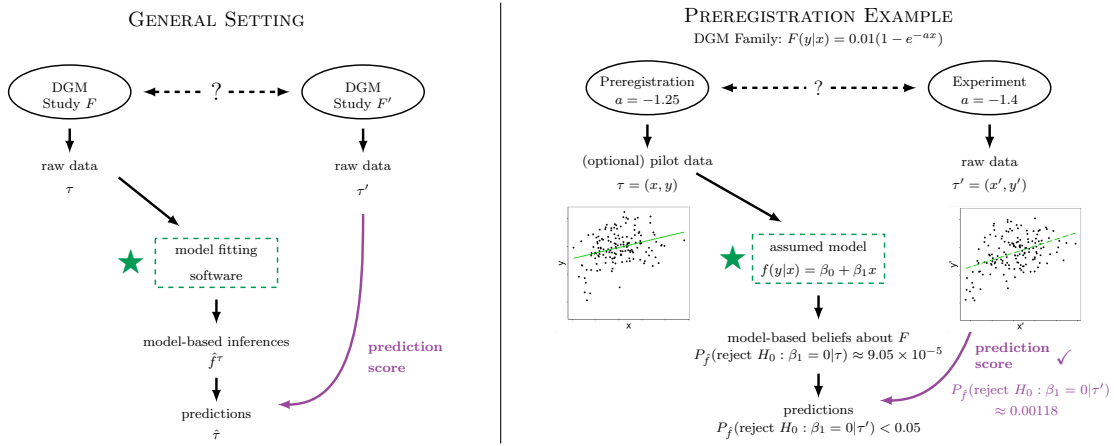
\begin{figure}
\begin{center}
\scalebox{.65}{
\begin{tikzpicture}[y=-1cm,scale=1]
\footnotesize
	
	\node at (5,-1) {\textsc{\large General Setting}};

	\draw[line width=.3mm] (2,0.5) ellipse (1.25cm and .55cm);
	\node at (2,0.4) {DGM};
	\node at (2,0.75) {Study $F$};
	
	\draw[latex-,dashed, line width=.5mm] (3.5,0.5) -- (4.75,0.5);
	\node at (5,0.5) {\Large ?};
	\draw[-latex,dashed, line width=.5mm] (5.25,0.5) -- (6.5,0.5);

	\draw[line width=.3mm] (8,0.5) ellipse (1.25cm and .55cm);
	\node at (8,0.4) {DGM};	
	\node at (8,0.75) {Study $F'$};
	
	\draw[-latex, line width=.5mm] (2,1.25) -- (2,1.75);
	\node at (2,2) {raw data};
	\node at (2,2.5) {$\tau$};
	
	\draw[-latex, line width=.5mm] (8,1.25) -- (8,1.75);
	\node at (8,2) {raw data};
	\node at (8,2.5) {$\tau'$};	
	
	\draw[-latex, line width=.5mm] (3,2.25) -- (4.25,3.25);
	\draw[ForestGreen, dashed, line width=.3mm] (3.75,3.5) rectangle (6.25,4.5);
	\node at (5,3.75) {model fitting};
	\node at (5,4.25) {software};
	\node[ForestGreen] at (3.25,4) {\LARGE $\bigstar$};
	
	\draw[-latex, line width=.5mm] (5,4.75) -- (5,5.25);
	\node at (5,5.6) {model-based inferences};
	\node at (5,6.0) {$\hat{f}^{\tau}$};
	\draw[-latex, line width=.5mm] (5,6.35) -- (5,6.85);
	\node at (5,7.1) {predictions};
	\node at (5,7.5) {$\hat{\tau}$};
	
	\draw[-latex,line width=.5mm,Purple] (8,3) to [out=270,in=0] (6.25,6.9);
	\node[below right, Purple] at (8,5.5) {\textbf{prediction}};
	\node[below right, Purple] at (8,6) {\textbf{score}};
\end{tikzpicture}}
\qquad\vrule\qquad
\scalebox{.65}{
\begin{tikzpicture}[y=-1cm,scale=1]
\footnotesize
	
	\node at (5.5,-1) {\textsc{\large Preregistration Example}};
	\node at (5.5,-.5) {DGM Family:  $F(y|x) = 0.01(1-e^{-ax})$};

	\draw[line width=.3mm] (2,0.5) ellipse (1.25cm and .55cm);
	\node at (2,0.4) {Preregistration};
	\node at (2,0.75) {$a = -1.25$};
	
	\draw[latex-,dashed, line width=.5mm] (3.5,0.5) -- (5.25,0.5);
	\node at (5.5,0.5) {\large ?};
	\draw[-latex,dashed, line width=.5mm] (5.75,0.5) -- (7.5,0.5);

	\draw[line width=.3mm] (9,0.5) ellipse (1.25cm and .55cm);
	\node at (9,0.4) {Experiment};	
	\node at (9,0.75) {$a=-1.4$};
	
	\draw[-latex, line width=.5mm] (2,1.25) -- (2,1.75);
	\node at (2,2) {(optional) pilot data};
	\node at (2,2.5) {$\tau = (x,y)$};
	
	\node at (1.5,4.2) {\includegraphics[width=.18\textwidth]{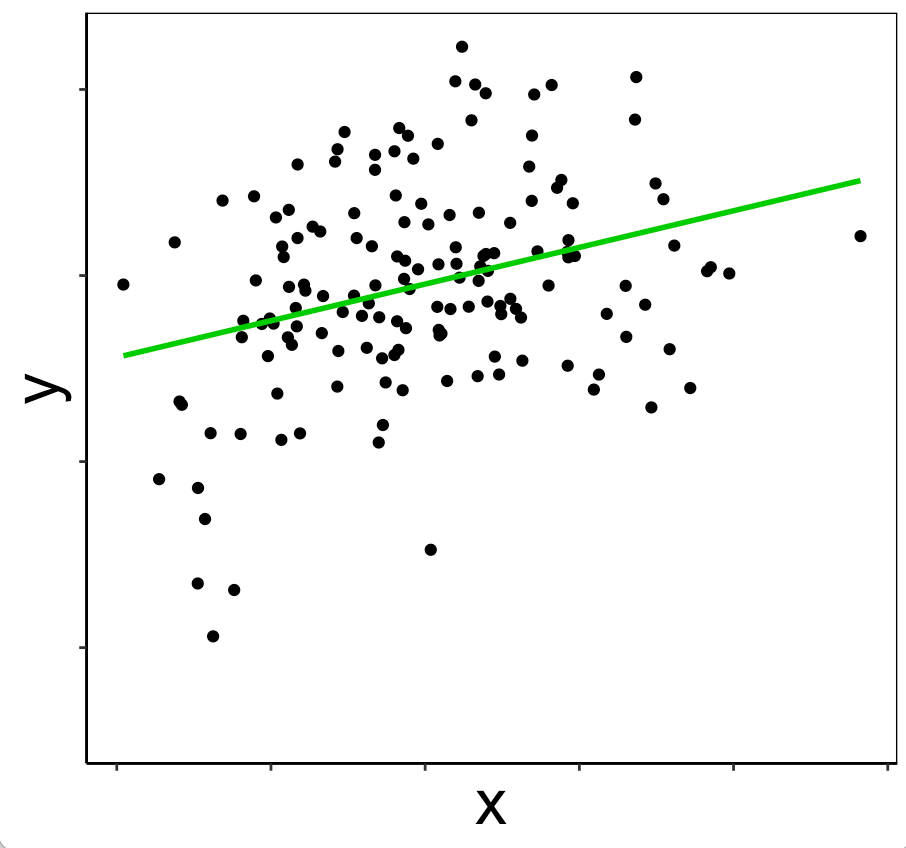}};
	
	\draw[-latex, line width=.5mm] (9,1.25) -- (9,1.75);
	\node at (9,2) {raw data};
	\node at (9,2.5) {$\tau'= (x',y')$};	
	
	\draw[-latex, line width=.5mm] (3,2.25) -- (4.75,3.25);
	\draw[ForestGreen, dashed, line width=.3mm] (4.15,3.5) rectangle (6.85,4.5);
	\node at (5.5,3.75) {assumed model};
	\node at (5.5,4.25) {$f(y|x) = \beta_0 + \beta_1 x$};
	\node[ForestGreen] at (3.75,4) {\LARGE $\bigstar$};
	
	\draw[-latex, line width=.5mm] (5.5,4.75) -- (5.5,5.25);
	\node at (5.5,5.6) {model-based beliefs about $F$};
	\node at (5.5,6.0) {$P_{\hat{f}}(\text{reject } H_0:\beta_1=0 | \tau) \approx 9.05 \times 10^{-5}$};
	\draw[-latex, line width=.5mm] (5.5,6.35) -- (5.5,6.85);
	\node at (5.5,7.1) {predictions};
	\node at (5.5,7.5) {$P_{\hat{f}}(\text{reject } H_0:\beta_1=0 | \tau') < 0.05$};
	
	\draw[-latex,line width=.5mm,Purple] (9,3) to [out=270,in=0] (6.75,6.9);
	\node[below right, Purple] at (8.6,5.6) {\textbf{prediction}};
	\node[below right, Purple] at (8.6,6) {\textbf{score}};
	\node[below right, Purple] at (10.25,5.75) {\large \checkmark};
	\node[below right, Purple] at (8.1,6.45) { $P_{\hat{f}}(\text{reject } H_0:\beta_1=0 | \tau')$};
	\node[below right, Purple] at (9,6.9) { $\approx 0.00118$};
	
	\node at (9.5,4.2) {\includegraphics[width=.18\textwidth]{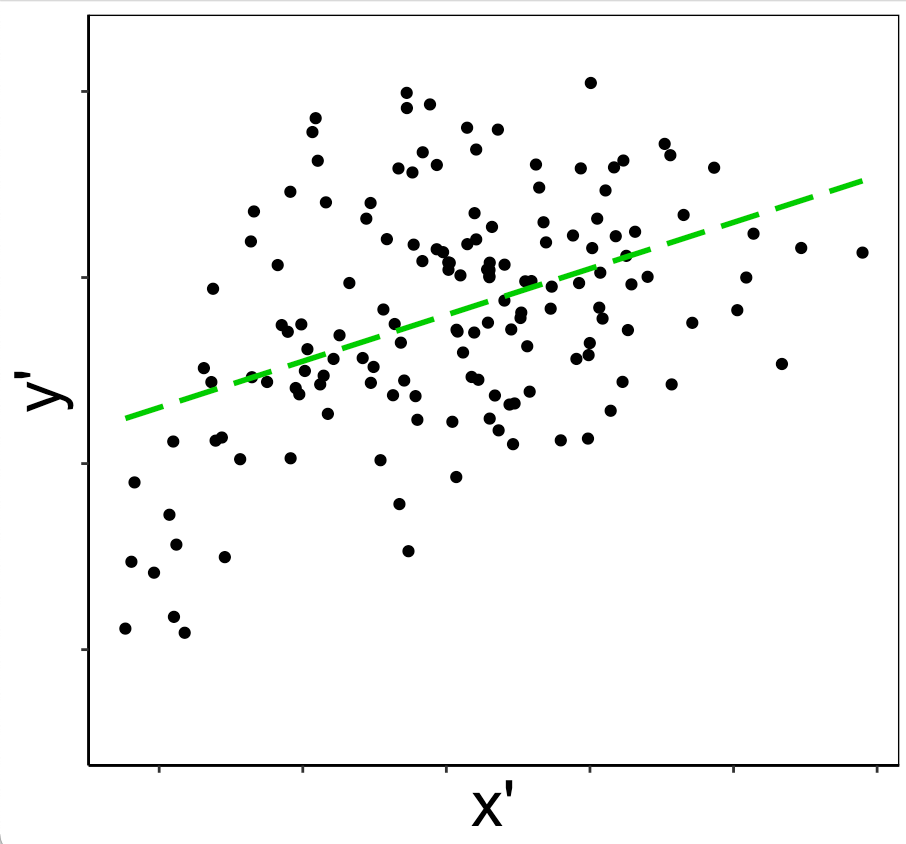}};
\end{tikzpicture}}

\centering
  \caption{Prediction scores are meant to measure the agreement between predictions and realized data.  For preregistered studies (right panel), common practice is to specify predictions in the form of null hypothesis statistical tests about model parameters; these predictions are then `scored' on a binary scale.  We formalize (and broaden) the concept of prediction scores as model-based continuous functions of predicted observables for the realized experimental data, $\tau'$, obtained from the assumed predictive model trained on some pilot or prior data, $\tau$ (left panel).}
  \label{fig:flowIntro}
\end{center}
\end{figure}

Returning to our motivating example of preregistration, consider the setting in which we have access to some pilot data or prior study that informs our preregistered hypotheses.\footnote{This is the case we will assume throughout the rest of the paper; if this is not the case, we might considering simulating some potential pilot data that incorporates whatever scientific beliefs we hold about the process and wish to study in our preregistered hypotheses}
Then, at the conclusion of a preregistered experiment, we are faced with two data sets: $\tau$, from the pilot study, and $\tau'$, the realized experimental data.
We assume that each of these datasets is generated from some unobservable data generating process---$F$ and $F'$---and we are interested in learning about underlying differences between $F$ and $F'$ (a definition for DGMs is discussed in Section~\ref{sec.cvPredScore}).
We propose evaluating the difference between (1) the set of (preregistered) predictions, $\hat{\tau}'$ and (2) the observed experimental data, $\tau'$, from $F'$.
These predictions, $\hat{\tau}'$, are predictions for the experimental data, $\tau'$, and are obtained from $\hat{f}^{\tau}$, a model for $F$ that is trained on the pilot data $\tau$ (See Figure~\ref{fig:flowIntro}).
In this section, we discuss some important motivating ideas for how exactly to use this comparison, between $\hat{\tau}'$ and $\tau'$, to learn about the quantity of interest, the underlying distance between $F$ and $F'$.
This comparison needs to decompose the total error (i.e., the observed difference between these model-based predictions, $\hat{\tau}'$, and the observed experimental data, $\tau'$) into estimation error (from estimating $F$ by $\hat{f}$) and model differences, i.e., true differences between the underlying DGMs.  

\paragraph*{Why NHSTs are often insufficient} To better understand the nuances of this type of comparison, consider an example with a simple linear regression model with data sampled from each DGM (as in the right panel of Figure \ref{fig:flowIntro}).  Common practice is to predict the outcome of NHSTs for parameters in an assumed predictive model.
When we have access to pilot data, we could simply merge the datasets and construct indicator variables, allowing for varying intercepts or slopes across the two samples.
Hypothesis tests for the corresponding regression coefficients can nicely summarize these specific types of differences between the two DGMs (see also Figure \ref{fig:trad_analysis} for a full example with real data).
But for more general cases, evaluating differences between DGMs is not a straightforward task.
For example, in the simple example given in the right panel of Figure \ref{fig:flowIntro}, the NHST from a simple linear regression model is incapable of detecting the more nuanced true difference between $F$ and $F'$.
In practice, simple linear regression simply cannot sufficiently summarize the scientific process under consideration and, even when more complex statistical models can be used, relying on a single parameter or summary measure is typically unsatisfying.  

Instead, we propose using appropriate predictive models to formulate scores for the predictions from one DGM $F$ against observables from another, $F'$.  As in predictive inference, this has the advantage of providing results that are highly interpretable and of direct interest to substantive researchers while also allowing for direct validation \citep[in a way that is simply impossible for model parameters;][]{geisser_2017, billheimer_2019}.  Further, this allows us to incorporate highly complex statistical models in exploring the unobserved DGMs, beyond linear regression models, and to move beyond effects described by a single parameter.  As in posterior predictive checking for Bayesian models, this choice allows for great flexibility in the types of differences that can be uncovered between the underlying DGMs.

\paragraph*{Controlling for model fit issues} In some ways, prediction scoring is similar to measuring predictive accuracy, but the goal here is different.  Using the notation in Figure \ref{fig:flowIntro}, both predictive accuracy and prediction scoring focus on discrepancies between $\hat{\tau}'$ and $\tau'$ but predictive accuracy uses these discrepancies to evaluate how well a model, $\hat{f}^{\tau}$, matches the true underlying DGM, $F$ (and often which of among a set of competing $\hat{f}^{\tau}$s provides the best match).  In this setting, it is typically assumed that $F = F'$; comparisons of this type are often called validation studies. In the prediction scoring setting, we do not assume that $F = F'$.  As a result, typical predictive accuracy metrics are model-dependent measures that mix together the discrepancies between the modeling framework and the underlying DGM family (i.e., model fit issues) {\em and} any differences between the DGMs. 

For this reason, we propose comparing validation summary measures to similar measures obtained from cross validation.  The resulting prediction scores are then adjusted for cross validation's estimate of how well the model fits the data.  Additionally, unlike other measures of model fit, cross validation is a general procedure that can accommodate many modeling frameworks \citep[although appropriate partitioning can be difficult for dependent or hierarchical data;][]{racine_2000,gelman_2006,roberts_bahn_ciuti_etal_2017} and is a clear analogue for the traditional validation procedure.

\paragraph*{The role of the predictive model} While cross validation can help to ``normalize'' or adjust for some of the effects of poor modeling choices, the chosen predictive model does impact the resulting prediction scores.  As a result, the predictive model acts as a lens through which we can view differences between the two underlying DGMs, which themselves cannot be directly observed.  Naturally, different models offer different perspectives and in the application described in Section \ref{sec.SimStudy} we leverage this aspect of our prediction scoring methodology by examining suites of non-nested predictive models to uncover distinct types of differences between the underlying DGMs.

\begin{figure}
\centering
\begin{tikzpicture}[y=-1cm]
\footnotesize
	
	\draw[line width=.3mm] (4.625,-2.5) ellipse (1.1cm and .75cm);
	\node at (4.625,-2.75) {DGM};
	\node at (4.625,-2.25) {family};
	\draw[-latex, line width=.5mm] (3.5,-2) to (2.5,-1);
	\draw[-latex, line width=.5mm] (5.75,-2) to (7,-1);

	\draw[line width=.3mm] (2,0) ellipse (1.5cm and .75cm);
	\node at (2,-0.25) {Experiment 1};
	\node at (2,0.25) {$F$};
	
	\draw[latex-latex,Purple,line width=.5mm] (3.65,0) -- (5.6,0);
	\node[Purple] at (4.625,-0.4) {$\Delta_{DGM}$}; 

	\draw[line width=.3mm] (7.25,0) ellipse (1.5cm and .75cm);
	\node at (7.25,-0.25) {Experiment 2};	
	\node at (7.25,0.25) {$F'$};
	
	\draw[-latex, line width=.5mm] (2,1) -- (2,2);
	\node at (2,2.5) {Data, $\tau$};
	
	\draw[-latex, line width=.5mm] (7.25,1) -- (7.25,2);
	\node at (7.25,2.5) {Data, $\tau'$};
		
	\draw[-latex, line width=.5mm] (2,3) -- (2,4);
	\draw[ForestGreen,dashed,line width=.5mm] (0.5,4.25) rectangle (3.5,5.5);
	\node at (2,4.7) {predictive model};
	\node at (2,5.1) {$\hat{f}^{\tau}(X)$};
	\node[ForestGreen] at (0,4.9) {\LARGE $\bigstar$};

	\draw[-latex,OrangeRed,dotted,line width=.5mm] (1.25,2.5) -- (0,2.5);
	\node[OrangeRed] at (-.75,2.5) {subsets};
	\node[OrangeRed] at (-.75,2.75) {of $\tau$};
	\draw[-latex,OrangeRed,line width = .5mm] (0,3) -- (1.5,4);
	
	\draw[-latex,NavyBlue,dotted,line width=.5mm] (8,2.5) -- (9.25,2.5);
	\node[NavyBlue] at (10,2.5) {subsets};
	\node[NavyBlue] at (10,2.75) {of $\tau'$};
	
	\draw[-latex,OrangeRed,line width = .5mm] (1.75,5.75) -- (1.75,6.8);
	\node[OrangeRed] at (1.75, 7.15) {predictions};

 	\draw[OrangeRed,dashdotted,line width = .5mm] (1.75,7.5) -- (1.75,8.025);
 	\draw[OrangeRed,dashdotted,line width = .5mm] (1.75,8) -- (-1.775,8);
 	\draw[OrangeRed,dashdotted,line width = .5mm] (-1.75,8) -- (-1.75, 1.75);	
 	\draw[OrangeRed,dashdotted,line width = .5mm] (-1.75,1.75) -- (-.75,1.75);	
 	\draw[-latex,OrangeRed,dashdotted,line width = .5mm] (-.75,1.75) -- (-.75,2.15);
 	\node[OrangeRed] at (0,8.5) {\textbf{cross validation}};
 	\node[OrangeRed] at (0,8.9) {loss statistics};
	

	\draw[NavyBlue,line width = .5mm] (2.25,5.75) --(2.25,6.15);	
	\draw[NavyBlue,line width = .5mm] (2.25,6.125) --(7.275,6.125);
	\draw[-latex, NavyBlue,line width = .5mm] (7.25,6.1) -- (7.25,6.8);
	\node[NavyBlue] at (7.25,7.15) {predictions};
	
 	\draw[NavyBlue,dashdotted,line width = .5mm] (7.25,7.5) -- (7.25,8.025);
 	\draw[NavyBlue,dashdotted,line width = .5mm] (7.25,8) -- (11,8);
 	\draw[NavyBlue,dashdotted,line width = .5mm] (11,8) -- (11,1.75);	
 	\draw[NavyBlue,dashdotted,line width = .5mm] (11,1.75) -- (10,1.75);	
 	\draw[-latex,NavyBlue,dashdotted,line width = .5mm] (10,1.75) -- (10,2.15);
 	\node[NavyBlue] at (9.125,8.5) {\textbf{validation}};
 	\node[NavyBlue] at (9.125,8.9) {loss statistics};
	
	
	\draw[latex-latex,Purple,line width = .5mm] (1.5,8.65) -- (7.5,8.65);
	\node[Purple] at (4.625,9.1) {$\Delta_{pred}$};
	
	\draw (-1.75,11.2) rectangle (11,10.2);
	\node at (-.5,10.5) {\textbf{practical}};
	\node at (-.5,10.9) {\textbf{considerations}};
	\draw (.75,11.2) -- (.75,10.2);
	
	\draw[ForestGreen,dashed,line width=.5mm] (1.2,10.5) rectangle (1.9,10.9);
	\node at (3.02,10.5) {predictive};
	\node at (3.02,10.9) {model};

	\draw[dashdotted,line width=.5mm] (4.27,10.7) -- (5.27,10.7);
	\node at (6.39,10.5) {loss};
	\node at (6.39,10.9) {statistic};
	
	\draw[dotted,line width=.5mm] (7.64,10.7) -- (8.64,10.7);
	\node at (9.76,10.5) {subset};
	\node at (9.76,10.9) {sampling};
	
\end{tikzpicture}
  \caption{General outline of the proposed prediction scoring methodology for generic data generating mechanisms, $F$ and $F'$.  We observe datasets $\tau = \left\{ (x_1, y_1), \dots (x_N,y_N)\right\}$ with data points drawn from $F$ and $\tau'= \left\{ (x_1', y_1'), \dots (x_N',y_N')\right\}$ with data points drawn from $F'$.  We believe that the target variable $y$ is related to a vector of inputs $X$ and estimate a predictive model, $\hat{f}^{\tau}(X)$, based on $\tau$.}
  \label{fig:flowAb}
\end{figure}
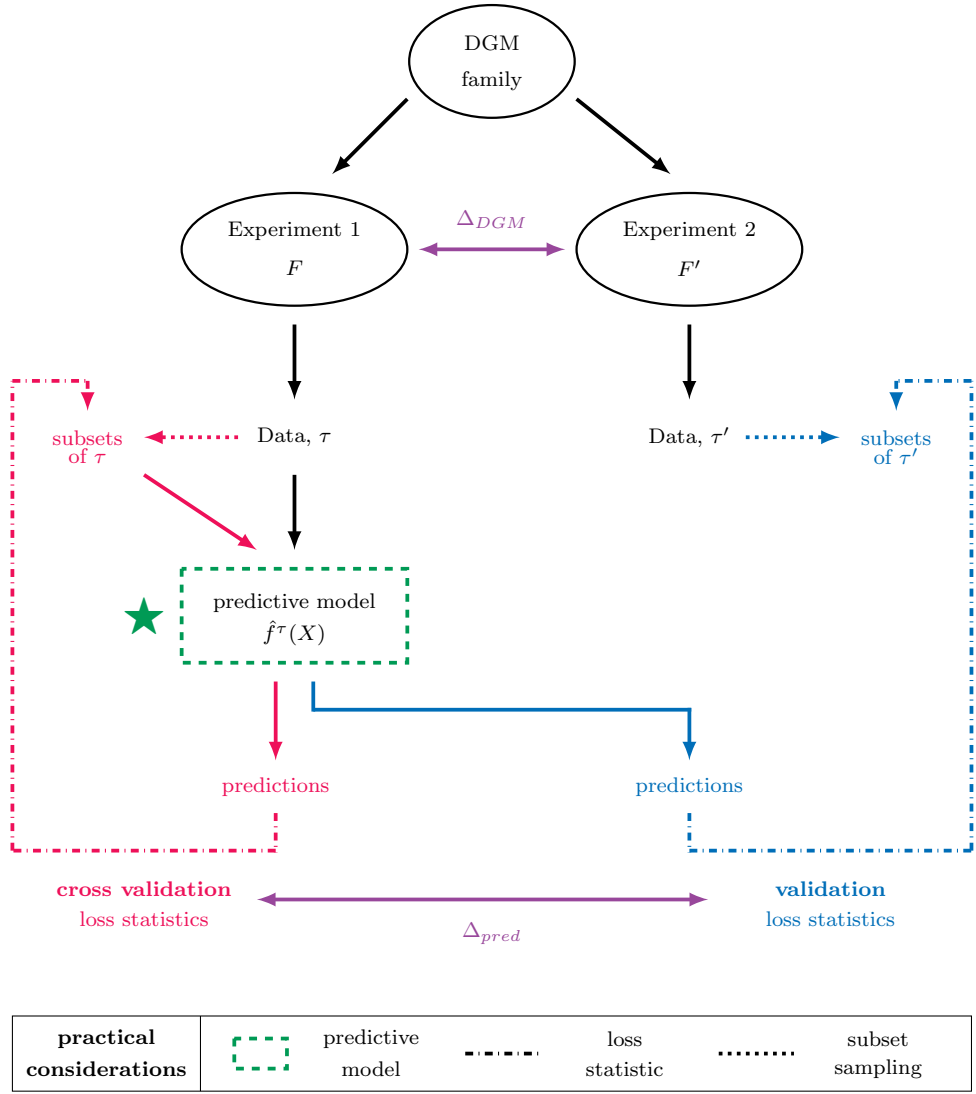

\paragraph*{Our proposed prediction scores} We provide full details of our proposed prediction scoring methodology in Section \ref{sec.cvPredScore}. 
In short, the idea is to learn about the distance between $F$ and $F'$ by comparing between model-based (preregistered) predictions, $\hat{\tau}'$, to observed experimental data, $\tau'$, using flexible loss functions to summarize the comparison and cross validation (along with subsampling of $\tau'$) to adjust for error due to model fit (see Figure \ref{fig:flowAb}; a more detailed discussion of this figure is given in Section \ref{sec.cvPredScore}).
In Section \ref{sec.SimStudy}, we consider differences across settings of a single simulated experimental setup and examine the probabilistic behavior of our proposed scores across many repetitions of the experiment.
In Section \ref{sec.AppNGS2}, we demonstrate how prediction scoring can be used to evaluate preregistered predictions from a human behavior experiment.
We discuss directions for future work in Section \ref{sec.Disc}.
Related methods for assessing predictive accuracy in the model selection setting and review recent advances in cross validation approaches are discussed in the appendix.

\section{Cross-validated prediction scoring} \label{sec.cvPredScore}

In our approach, we will use the term \textit{data generating mechanism} to refer to the unobservable underlying stochastic process that describes the scientific phenomenon under study.  To be precise, we will conceptualize a DGM as a particular member of a family of probability distributions that represent a set of (model) assumptions about the scientific phenomenon 

For any family of data generating mechanisms, we are interested in estimating a distance, or measure of dissimilarity, between different members of the same family.

\vspace{0.3cm}
\begin{definition}
\label{def:dist}
For a particular family of data generating mechanisms, let the dissimilarity between any two members of the family be represented by
\[ \Delta_{DGM} = d \left( F, F'\right) \]
where $F$ and $F'$ are both members of a particular DGM family and the choice of the dissimilarity function $d$ is motivated by the form of the DGM family.
\end{definition}

We will focus on regression-like settings where we are interested in learning about the relationship between a target variable, $y$, and a vector of inputs, $X$.  As a result, we will think of the underlying DGM $F$ as representing the joint population distribution for the data, but that research interests are focused on learning the conditional relationship, $p(y|X)$.  This is a natural framework since, for example in a human behavior study, we may be interested in learning about the conditional relationship between individuals' demographic characteristics and some behavioral outcome but believe that a joint distribution describes the way that the sampled data are drawn from the population of all possible studies (i.e., for a new sample, we would observe different individuals with different demographics and different behavioral outcomes).

\subsection{Predictive accuracy and test error}

To motivate the form of the prediction scores proposed here, we first review classical definitions of predictive error.  As mentioned previously, the prediction scores essentially compare model-based predictions to real-world observations (see Figure \ref{fig:flowIntro}), and so traditional model assessment tools are useful.  In the model assessment setting, we are interested in evaluating the performance of a final selected model by estimating the model's predictive error \textit{for new data}.  Let $\hat{f}^{\tau}(X)$ be a predictive model estimated from a fixed dataset $\tau = \left( (x_1, y_1), \dots, (x_N,y_N) \right)$ drawn from $F$.  For an appropriate loss function, $L(y,\hat{f}^{\tau}(X))$, model assessment tools typically estimate the conditional test error \citep[also called the prediction or generalization error;][]{hastie_tibshirani_friedman_2001}, which depends on the particular (fixed) training set $\tau$:
\[ \text{Err}_{F|\tau} = \text{E}_{X,y \sim F} \left( L\left( y,\hat{f}^{\tau}(X) \right) | \tau \right). \]
This is the expected error for the predictive model trained on the dataset, $\tau$, and the expectation is over all new data drawn from $F$ but is conditional on observing the particular training data in $\tau$.  In practice, a new dataset is typically not available, and so this error is estimated using only the original observations, for example by performing cross validation (or bootstrapping or by calculating AIC, BIC, etc.) which estimates the expected test error, $\text{E}_{\tau}[\text{Err}_{F|\tau}]$, an average of the (conditional) test error over all possible training sets $\tau$.

In the prediction scoring setting, we are considering two data generating mechanisms, or population distributions, $F$ and $F'$, each of which are joint distributions for the data.  From each of these distributions, we observe a dataset, $\tau$ and $\tau'$ respectively.  In the model assessment setting we care about the difference between $F$ and $\hat{f}$ (i.e., how well does the model estimate the truth), whereas in the prediction scoring setting we care only about differences between $F$ and $F'$.  Since we cannot observe the DGMs directly, we can use a predictive model to summarize differences in the observed data; in this sense, the predictive model is like a nuisance parameter that we cannot avoid since the DGMs themselves cannot be directly observed (we only observe the datasets $\tau$ and $\tau'$).

Alternatively, we could consider measuring differences between the DGMs through explicit differences across the datasets themselves.  For example, consider the test statistic for the two-sample $t$-test which is a function of the sample means.  This test (and others like it) assume an underlying parametric model; the test is designed to detect differences between parameters from this model.  

A natural model-based approach is to perform validation, where the predictive model is trained on (some subset of) the first (training) dataset, $\tau$, but evaluated in the context of the new (test) data, $\tau'$.  The estimate of predictive error given by validation is typically of the following form,
\[ \text{Val}( \hat{f}^{\tau} ) = \frac{1}{N'} \sum_{i=1}^{N'} L \left( y_i' , \hat{f}^{\tau} \left(x_i' \right) \right) \]
where $N'$ is the number of observations in $\tau'$.  In this sense, validation error estimates a different conditional test error, given by
\begin{eqnarray*}
\text{Err}_{F'|\tau} = \text{E}_{X',y' \sim F'} \left( L\left( y', \hat{f}^{\tau}(X') \right) | \tau \right),
\end{eqnarray*}
where here the expectation averages over draws from $F'$. 

However, model validation captures differences due to \textit{both} model fit issues (from estimating $F$ by $\hat{f}^{\tau}$) and true differences between the DGMs (between $F$ and $F'$).  Instead of relying solely on validation measures, we propose using cross validation to properly calibrate the measurements from validation.  In this way, we can separate the differences due to model fit issues and random variation (as measured by cross validation) from any true differences between the data generating mechanisms.  

\subsection{General framework} \label{subsec.GenFrame}
Letting $\kappa : \left\{ 1, \dots, N \right\} \rightarrow \left\{ 1, \dots, K \right\}$ be an indexing function specifying data splits, the estimate of predictive error given by $K$-fold cross validation is
\[ \text{CV}_{\kappa} ( \hat{f}^{\tau} ) = \frac{1}{N} \sum_{i=1}^N L \left( y_i, \hat{f}^{\tau}_{-\kappa(i)} \left(x_i \right) \right), \]
where $\kappa$ is typically specified such that the number of observations in each of the $K$ partitions is roughly equal.  Cross validation estimates the \textit{expected} test error, $\text{E}_{\tau}[\text{Err}_{F|\tau}]$.

Our prediction scores are designed to compare differences between validation and cross validation.  In order to make this comparison meaningful, we need to consider a version of validation that estimates the conditional test error, $\text{Err}_{F'|\tau}$, averaged over new potential draws of $\tau$ from $F$:  $\text{E}_{\tau}[\text{Err}_{F'|\tau}]$.  This can be achieved by redefining the typical validation loss statistics as follows:
\[ \text{Val}_{\kappa}( \hat{f}^{\tau} ) = \frac{1}{N'} \sum_{i=1}^{N'} L \left( y_i' , \hat{f}^{\tau}_{-\kappa(i)} \left(x_i' \right) \right) \]
where the $\hat{f}^{\tau}_{-\kappa(i)}$'s are the \textit{same} predictive models from cross validation (i.e., the models are trained on the same \textit{subsets} of $\tau$).  This differs from the traditional implementation of validation in which the predictive model would be trained using all entries in $\tau$.  Intuitively, if this were the case, we would naturally expect better predictive performance in the validation routine, since the predictive model has the benefit of being trained on more data.  Keeping the predictive models as comparable as possible (across the cross validation and validation routines) by training them on the same subsets of the data enables better detection of true differences between the underlying DGMs.

Differences between our redefined validation and cross validation estimate differences between the conditional test errors based on $F$ and $F'$, averaged over new training datasets $\tau$.  In other words,
\[ \text{Val}_{\kappa}(\hat{f}^{\tau}) - \text{CV}_{\kappa}(\hat{f}^{\tau}) \text{ estimates } \text{E}_{\tau}\left[ \text{Err}_{F'|\tau} \right] - \text{E}_{\tau}\left[ \text{Err}_{F|\tau} \right] \]
and the estimand can be expanded as
\[ \text{E}_{\tau} \left\{ \int_{\mathcal{X,Y}} L\left( y, \hat{f}^{\tau}(x) \right) d \left( F(x,y|\tau) - F'(x,y|\tau) \right\} \right), \]
which is the prediction error or loss averaged over differences between the DGMs and averaged over training sets $\tau$.  This expansion elucidates a natural connection; the differential resembles the form of the Kolmogorov-Smirnov statistic, a popular method that uses differences in empirical cumulative distribution functions (i.e., estimates for $F$ and $F'$) to measure discrepancies across univariate distributions.

\begin{algorithm}[t]
\caption{Prediction Scoring}\label{alg}
\begin{flushleft}
Let $\tau = \left\{ (x_1, y_1), \dots (x_N, y_N) \right\}$ be a dataset with $N$ observations drawn from the DGM $F$ which is the joint population distribution for $X$ and $y$; and analogously, $\tau'$ consists of $N'$ observations and is drawn from $F'$.  Let $\kappa: \left\{1,\dots,N \right\} \rightarrow \left\{ 1, \dots, K \right\}$ be an indexing function specifying data splits for $\tau$ and similarly $\kappa': \left\{1,\dots,N' \right\} \rightarrow \left\{ 1, \dots, K \right\}$ specifies data splits for $\tau'$.  Let $L$ be an appropriate loss function and let $h$ be the prediction scoring function.
\end{flushleft}
\begin{algorithmic}[1]
\Procedure{Fit Predictive Models}{}
\For {$k=1, \dots K$}
\State \text{construct $\tau_{-k} = \left\{ (x_i, y_i) \text{ s.t. } \kappa(i) \ne k \right\}$} 
\State \text{compute} $\hat{f}^{\tau}_{-k}$, \text{the predictive model trained on $\tau_{-k}$}
\EndFor
\EndProcedure
\Procedure{cross validation}{}
\For {$i=1, \dots N$}
\State $l_i = L \left( y_i, \hat{f}^{\tau}_{-\kappa(i)}(x_i) \right)$
\EndFor
\EndProcedure
\Procedure{Validation}{}
\For {$i=1, \dots N'$}
\State $l_i' = L \left( y_i', \hat{f}^{\tau}_{-\kappa'(i)}(x_i') \right)$
\EndFor
\EndProcedure
\Procedure{Prediction Scoring}{}
\State $\Delta_{pred} = h \left( l, l' \right) $
\EndProcedure
\end{algorithmic} \label{alg.predscore}
\end{algorithm}

In Algorithm \ref{alg} and Figure \ref{fig:flowAb}, we generalize the prediction scores so that comparisons between the cross validation and validation summary measures need not be computed only as the difference of means, as above.  That is, our prediction scores are defined as
\[ \Delta_{pred} = h \left( l, l' \right), \]
where $l$ and $l'$ are the vectors of loss statistics such that $l_i = L \left( y_i, \hat{f}^{\tau}(x_i) \right)$ for $i=1, \dots, N$ and $l_i' = L \left( y_i', \hat{f}^{\tau}(x_i') \right)$ for $i=1, \dots, N'$.
Although in our definition above the prediction score is a \textit{function} that compares the loss statistics, in practice, diagnostic plots that represent the differences between these distributions can be useful, as demonstrated in later examples.

\subsection{Additional considerations}

\paragraph*{Forecast distributions and non-Bayesian models} In order to appropriately account for uncertainty, we recommend making predictions in the form of vectors of possible outcomes, also called simulated forecast distributions. In the examples that follow we will adopt Bayesian predictive models which provide a natural way of computing vector-valued predictions; we can simply draw samples from the posterior predictive distribution for each observation.  For non-Bayesian models, similar types of predictive distributions can be computed with bootstrapping or other resampling methods.

\paragraph*{Choosing appropriate functionals} In practice, calculating prediction scores involves specifying a few important elements:  the predictive model ($\hat{f}$), the loss function to compare predictions to realized data ($L$), and the subsampling method for the cross-validation and validation routines ($\kappa$).  Additionally, to study the theoretical properties of the prediction scores, appropriate choices for $d$ (the measure of the true difference between the data generating mechanisms) and $h$ (the measure of the difference between the distributions of the loss statistics) must be made.  These choices should be well motivated by the data types and modeling choices of the particular application.  More specifically, the true DGM distance should be motivated by the form of the family of data generating mechanisms being considered, and the loss function should be motivated by the form of the chosen predictive model and model fitting software.

Because we encourage making predictions in the form of forecast distributions, the loss function needs to be capable of evaluating differences between the true observation, $y_i$, (a number) and the corresponding set of predictions (a vector).  This still leaves questions as to the form of the loss in the face of different types of predictive models.  For example, when using a linear regression model, predictions for $y$ will be continuous and so quantiles may be a natural choice.  However, for a logistic regression model, the predictions will be probabilities (between 0 and 1) while the observations are binary.  Some variant of the area under the curve (AUC) statistic may be a better choice for $L$.  Strictly proper scoring rules such as the logarithmic score or Brier score could be easily incorporated \citep{gneiting_raftery_2007}.

Additionally, the choice of the prediction score, $h$, should be motivated by both of these considerations and the subsequent choices for $d$ and $L$.  Although this methodology would be simpler if $d$, $L$, and $h$ were universally specified, it is important that they capture relevant features of the data generating mechanisms and are suitable to whatever modeling assumptions and model fitting software is chosen by the researcher.  Further, this sort of conditional specification is not unlike the choice of an appropriate link function for generalized linear models. This framework is nicely aligned with many popular measures of predictive accuracy.  For example, choosing $L$ as quadratic loss for a linear regression predictive model and choosing $h$ appropriately will result in prediction scores that compare mean squared error across the cross validation and validation routines.  In the examples discussed here we consider logistic regression predictive models and adopt the popular AUC statistic for $L$, examining histograms of these statistics in Section \ref{sec.AppNGS2} and computing $h$ as Kolmogorov-Smirnov statistics in Section \ref{sec.SimStudy}.  Deriving appropriate forms of $d$, $L$, and $h$ for more dependent data, such as networks or time series, is an active area of future research.  Ideally, the prediction scoring methodology, including these choices for $d$, $L$, and $h$, should be fully preregistered prior to any data collection.  This does not preclude us from using the prediction scores \textit{in an exploratory} fashion to discover interesting data features.

\subsection{The predictive model as a lens}

\begin{figure}
\centering
    \includegraphics[width=0.95\linewidth]{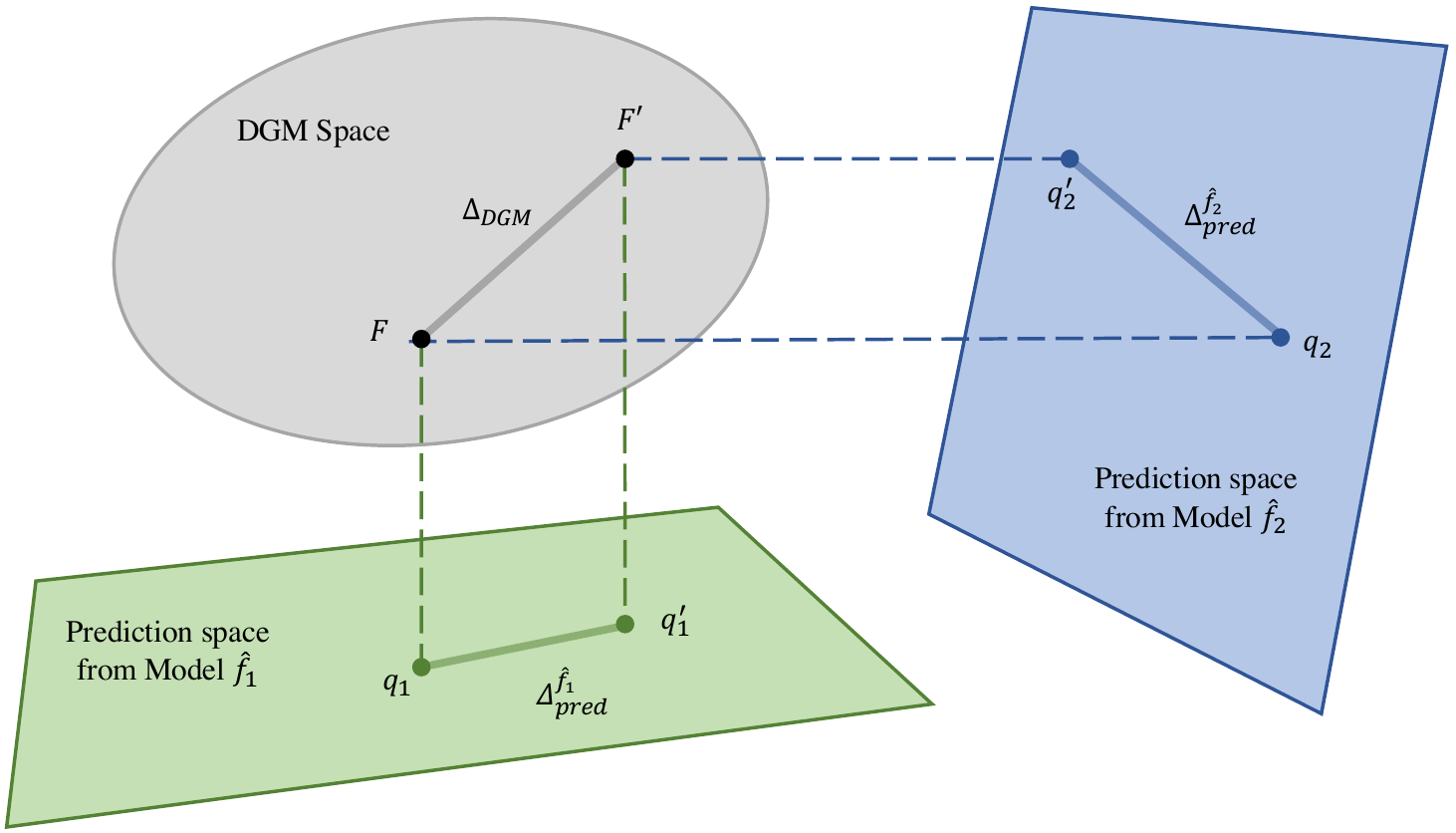}
  \caption{A geometric illustration of how prediction scores measure differences between DGMs \textit{along the dimension} of the model used to make predictions.}
  \label{fig:geom}
\end{figure}

As we will emphasize in the following examples, the proposed prediction scores reveal differences between the DGMs along the dimension of the model used to make predictions.  Consider evaluating differences between the same pair of DGMs in the face of two competing models.  In the case where these predictive models are orthogonal in some sense (i.e., capture distinct features of the DGMs), we can imagine that each model should produce a set of prediction scores that capture differences between the DGMs only according to the features of the DGMs that each model is equipped to detect.  For example, consider the illustrative diagram given in Figure \ref{fig:geom}.  Here, we imagine two DGMs, $F$ and $F'$, which reside in a large, complex, multidimensional DGM space.  The true distance between these DGMs, $\Delta_{DGM}$, is typically unobservable, but we can calculate prediction scores relative to a model, which measure the distance between the distributions of cross validation and validation loss statistics.  For example, using model $\hat{f}_1$, we can learn about the DGMs, $F$ and $F'$, by projecting them into a lower-dimensional space, the prediction space for model $\hat{f}_1$ (represented by the low-dimensional green rectangle in Figure \ref{fig:geom}).  In this lower-dimensional space, we can measure the distance between the predictive accuracy of model $\hat{f}_1$ for data corresponding to $F$ (this is represented in the empirical distribution for $q_1$ and is obtained via cross validation) and for data corresponding to $F'$ (represented by $q_1'$, obtained via validation).  This prediction scoring distance, $\Delta_{pred}^{\hat{f}_1}$ depends on the model used to make predictions, $\hat{f}_1$, and will correspond to differences between $F$ and $F'$ that the model is equipped to detect.  Now if we imagine a competing model, $\hat{f}_2$ (its prediction space is represented by the low-dimensional blue rectangle in Figure \ref{fig:geom}), which is ``orthogonal'' to this model, the resulting prediction scoring distance for model $\hat{f}_2$ will reflect different types of differences between the DGMs (i.e., in Figure \ref{fig:geom}, the blue distance is not the same as the green distance).  In other words, DGMs may look more or less similar in terms of prediction scoring distances, depending on the model used to make the predictions.  As we will discuss in more detail in Section \ref{sec.SimStudy}, we can leverage this dependence and consider suites of ``orthogonal'' models in order to discover interesting differences between DGMs.

\section{Simulated experiments} \label{sec.SimStudy}

To demonstrate that these prediction scores can pick up on meaningful differences between competing DGMs (here, across experimental settings), we have designed a simulation study that utilizes a simplified experimental design, modeling the outcome of interest with logistic regression.  The details of the simulated experiments are motivated by experimental data from a large-scale human behavior study, the Next Generation Social Science (NGS2) program (more details are available in the Appendix; experimental data from this program is explicitly considered in Section \ref{sec.AppNGS2}).  This simulation study also allows us to examine how prediction scores can leverage competing predictive models to identify different types of differences between DGMs and to get a sense of their asymptotic behavior.


\subsection{Experimental setup: human behavior in the presence of bots}

Our simulation study will compare multiple settings of the simple and well-studied {\em public goods} game, from economic game theory \citep{ledyard_1995}. In a public goods game, $n$ players have the opportunity to contribute (``cooperate") or not (``defect'') over a series of $T$ sequential rounds to a set of pooled resources that will be (multiplied and) shared among all participants.  Each player's goal is to collect as many tokens (money) as possible; in each round, players are faced with the decision to be selfish (and keep their tokens), or be cooperative (and donate money to the common pool).  In our simulated experiments, we assume that each player's decisions are made public to all other participants.  Economists hypothesize that players' decisions to contribute at each round depends on the players' own baseline tendency to contribute, their previous decisions, and can be affected by the outcomes and behaviors of other players from the previous rounds.

Inspired by the experimental plan of the NGS2 research teams from the University of Pennsylvania and University of California, we imagine that bot-like participants play alongside the simulated human participants.  Theoretically, inserting bot participants within these experiments would allow researchers greater experimental control over the social and environmental landscape within the game \citep{penn_cycle2} while simultaneously enabling the study of human behaviors in larger groups (i.e., adding bot participants is easier than recruiting human subjects).  In this sense, researchers can use bot behaviors to create interventions and trigger different behaviors.

\paragraph*{True DGM} In our simulation study, we consider an array of $K=5$ different DGMs, representing different levels of the percent of bot participants in the game, from $\pi = \left\{ 0, 0.25, 0.50, 0.75, 1 \right\}$. We are most interested in understanding the ways in which participants' decisions to cooperate are influenced by the presence of bots;  thus, prediction scores will compare predictions for participant contribution across experimental settings where the percentage of bots differs.



For each experimental setting (i.e., each element of $\pi$), we imagine recruiting $J$ cohorts of individuals to participate; let $n_{jk}$ be the number of individuals competing in the $j$th cohort of the $k$th setting.  Let $y_{ijkt}$ be the decision to cooperate ($y_{ijkt}=1$) or defect ($y_{ijkt}=0$) for the $i$th individual in the $j$th cohort of the $k$th experimental setting during round $t$, where $i = 1, \dots n_{jk}, j=1, \dots J, k=1, \dots K,$ and $t = 1, \dots T$.  Additionally, let $z_{ijk}$ be an indicator of whether the $i$th participant in the $j$th cohort of the $k$th round is a human participant ($z_{ijk}=1$) or a bot ($z_{ijk}$).  We will assume that $y_{ijkt}|z_{ijk}$ are independently distributed Bernoulli random variables, for all $i,j,k,$ and $t$.  Then, the true underlying data generating mechanism for the simulated data in our hypothetical experiments is given by the following:
\begin{align*}
&&z_{ijk} & \overset{iid}{\sim} \text{Bernoulli}(\pi_k) \\
&\text{Model 0:} & \text{logit}^{-1} \left( P( y_{ijkt} = 1 | z_{ijk} = 1) \right) & = \beta_0 + \beta_1 t + \beta_2 y_{ijk,t-1} + \beta_{3} \bar{y}_{\cdot j k, t-1} \\
&&\text{logit}^{-1} \left( P( y_{ijkt} = 1 | z_{ijk} = 0) \right) & = \beta_0' + \beta_2' y_{ijk,t-1},
\end{align*}
where $\pi_k$ is the proportion of bots in the $k$th round, $\beta_0$ and $\beta_0'$ are baseline tendencies to cooperate, $\beta_1$ captures any trend across the rounds, $\beta_2$ and $\beta_2'$ capture the tendency to switch between behaviors, and $\beta_{3}$ represents the influence of team members' decisions.  For example, if all other individuals cooperated in the previous round ($\bar{y}_{\cdot j k, t-1}$ is close to one), then the probability that individual $i$ also cooperates in the next round is high, for large positive $\beta_{3}$.
For bot participants, $\beta_{3}$ is defined to be zero; the simplistic bots we consider here are not influenced by the behavior of other participants.  To specify reasonable parameter values for our simulation, we fit this true model to the experimental data from both \citet{rand_arbesman_christakis_2011} and \citet{gallup_cycle1} (analyzed in Section \ref{sec.AppNGS2}), using data from games played under the fluid network update setting for bot behavior and the fixed network setting for human behavior (see Table \ref{tab.simCoefs}).

\begin{table}
    \centering
    \begin{tabular}{ l  c c c c }
         Parameter & $\beta_0$ & $\beta_1$ & $\beta_2$ & $\beta_3$ \\
         Covariate & $1$ & $t$ & $y_{ijk,t-1}$ & $\bar{y}_{\cdot jk, t-1}$ \\
         \hline 
         Human behavior & $-1.31$ & $-0.10$ &  1.97 & 1.25\\
         Bot behavior & $-0.78$ & - &  2.68 & - 
    \end{tabular}
    \vskip1ex
    \caption{Parameter values for Model 0: True data generating mechanism.}
     \label{tab.simCoefs}
\end{table}

To mimic subject recruitment, for each setting, $k$, we will set the number of cohorts $J = 10$ and the number of rounds $T = 15$, and draw $n_{kj} \sim \text{Binomial} ( M, p )$, where $J$ and $T$ are chosen to mimic the experimental settings specified by \citet{rand_arbesman_christakis_2011} and \citet{gallup_cycle1}, $M=10000$ is the size of the pool of possible recruits, and $p=0.0018$ is the participation rate; this corresponds to roughly 18 participants per cohort.

\sloppy
\paragraph*{Prediction scoring} 

To mimic our analysis of experimental data in Section \ref{sec.AppNGS2}, we will fit Bayesian logistic regression models (this also aligns well with the true underlying DGMs).  As discussed in Section \ref{subsec.GenFrame}, the form of the predictive model can help to inform our choice of loss function.  Natural loss functions for these models include ROC or precision-recall curves.\footnote{Generally, the precision-recall curve is preferred over the ROC curve when data are imbalanced, for example when there are many more 0's than 1's \citep[see][for more discussion]{davis_goadrich_2006} In our simulated data, even across each setting (i.e., where we compare data with $\pi_{k_1}$ bot participants to data with $\pi_{k_2}$ bot participants) the aggregate baseline cooperation rate varies from 0.44 (when both $\pi_{k_1}$ and $\pi_{k_2}$ are close to 0) to 0.66 (when both $\pi_{k_1}$ and $\pi_{k_2}$ are close to 1; these differences in the baseline rates are also apparent in the upper left panel of Figure \ref{fig.rocAll}).}, and the corresponding area under the curve statistics.  We will consider visual comparisons of these curves as well as differences in the distributions of the AUC statistics.  Data subsets are created as random subsamples containing roughly $500$ observations each; since we are not investigating cohort- or individual-level effects of any kind, the data are partitioned completely randomly across all observations but is resampled in order to preserve consistent class proportions across all subsets.  This resampling method is necessary to ensure that the precision-recall curves are comparable across datasets that vary by baseline cooperation rates (this is discussed in more detail in Section \ref{sec.AppNGS2} and in Panel I of Figure \ref{fig:ps_analysis}).

\paragraph*{Researcher models}  We consider a suite of three potential researcher models: 
\begin{eqnarray*}
\text{Model 1:} & \text{logit}^{-1} \left[ P( y_{ijkt} = 1) \right] &= \gamma_0 + \gamma_1 t, \\
\text{Model 2:} & \text{logit}^{-1} \left[ P( y_{ijkt} = 1) \right] &= \gamma_0' + \gamma_2 y_{ijk, t-1}, \\
\text{Model 3:} & \text{logit}^{-1} \left[ P( y_{ijkt} = 1) \right] &= \gamma_0'' + \gamma_3  \bar{y}_{\cdot jk, t-1},
\end{eqnarray*}
where $\gamma_0$ is a baseline tendency to cooperate, $\gamma_1$ can capture some trends across the rounds, $\gamma_2$ represents the influence of the most recent decision, and $\gamma_3$ represents the influence of team members' decisions.  In practice the true data generating mechanism is unknown to the researcher.  However, the researcher typically has hypotheses about features of the DGMs that might differ across experimental settings and these features are incorporated in models as above.  For example, if all participants are bots, than Model 2 should perform fairly well.  However, whenever humans participate, Model 2 will fail to represent the full spectrum of observed behaviors well.  The models specified here are intentionally non-nested.  Since prediction scores are inherently model-based (i.e., they depend on the model used to make the predictions), recall from Figure \ref{fig:geom} that we can interpret them as a distance between DGMs \textit{along the dimension of a particular model}.  In this sense, when trying to uncover features of the DGM that may differ across settings, models that can measure distinct features of the DGM should be prioritized.  In some sense, we can think of the desired set of researcher models as being orthogonal to each other\footnote{Here we mean that the models should be non-nested, but we use the term ``orthogonal'' to better relate to the geometric description of prediction scores provided earlier---that they measure distance along the dimension of a particular model.} so as to maximize the possibility of discovering true differences between the DGMs.

\subsection{Results}

\begin{figure}
\includegraphics[width=\textwidth,page=1]{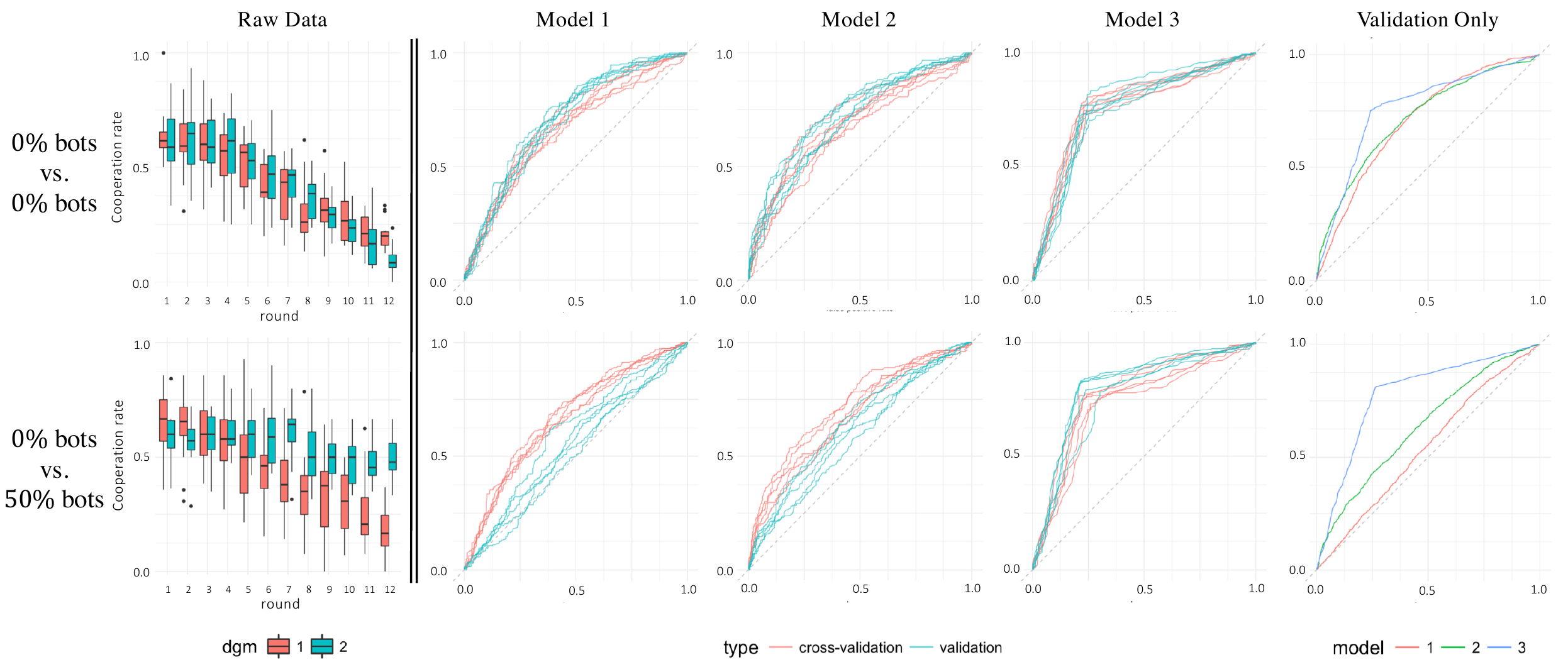}
\caption{ROC curves (true positive rate, $y$-axis, vs. false positive rate, $x$-axis) from prediction scoring to compare experimental settings.  Across the top row, experimental settings are identical (i.e., the underlying DGMs are identical and so $F = F'$) with all human participants ($\pi= 0$) while in the bottom row, the experimental settings differ (i.e.,$F \ne F'$) and compares all human participants ($\pi_1 = 0$) to 50\% bot participants ($\pi_2 = 0.50$).  The plots on the far left contain boxplots of the cooperation rate across all cohorts and individuals by round, where the color represents the experimental settings.  The plots on the far right are ROC curves for validation only, with each curve corresponding to a different researcher model.  Remaining plots display the prediction scoring ROC curves for subsets of the data from the cross validation (red) and validation (blue) routines, with each plot corresponding to a different researcher model.}
\label{fig.rocExample}
\end{figure}

\paragraph{Shortcomings of predictive accuracy}  First consider the more traditional approach of performing validation alone.  In this case, the posterior predictive distribution is conditioned on the \textit{full set of data} from the first experiment or dataset (covariates, $x$, and responses, $y$) but provides a prediction for the responses, $y'$, from the second experiment, corresponding to the covariates in that second experiment, $x'$.  This procedure is often used to compare competing models, such as those considered in our suite of researcher models here (see the far right panels in Figure \ref{fig.rocExample}).  Whether the underlying DGMs differ or not (in the top row, both DGMs have $\pi_1=\pi_2=0$ bots; the bottom row compares data from DGMs with $\pi_1$=0 and $\pi_2=0.50$), we observe that Model 3 has the best predictive accuracy.  However, using these curves alone, it is impossible to say much about any underlying differences between the DGMs being compared.  First, these curves fail to account for sampling variability.  If the sample of participants in either experiment differed slightly, we would expect to see the curves in these figures move around a bit, but just how much they would move (i.e., how much sampling variability for this particular population or experiment impacts model fit and predictive ability) can not be estimated or accounted for in the validation-only procedure.  In this sense, prediction scoring goes above and beyond traditional predictive accuracy measures; using subsets of the data in both the cross validation and validation routines helps to appropriately account for the effects of sampling variability.  Secondly, these curves do not allow us to separate the effects of (poor) model fit from any true differences between the DGMs.  Only by comparing cross validation curves to validation curves are we able to observe these differences.  Both cross validation and validation curves are based on predictions made from the same model, so that any observed differences should solely reflect true differences between the DGMs.

\paragraph{Detecting DGM differences} First, consider the case where the DGMs are in fact identical across settings; see the top panel of Figure \ref{fig.rocExample}.  As we would expect, there is little discernible difference between the cross validation and validation curves, regardless of the model used to make predictions.  If instead we consider the case where there is a difference between the DGMs, such as $\pi_1=0$ and $\pi_2=0.50$ as in the bottom panel of Figure \ref{fig.rocExample}, we can see some evidence of a difference between the DGMs as we would expect.  In short, the prediction scores are successful in detecting differences between the underlying DGMs.

\paragraph{Leveraging competing models}  Recall from Figure \ref{fig:geom}, that the prediction scores are dependent on the predictive model and measure differences between DGMs along the dimension of the model used to make predictions.  For Model 1, there is clear separation of the cross validation and validation curves indicating that there is a difference between these DGMs.  For Models 2 and 3, this difference is less clear, but there does appear to be an ordering of the curves which is some indication of a difference across the two DGMs.  Model 1 depends only on the round number.  In fact, if we look at the raw data simulated for these experiments we see strong differences over time across these two settings.  Thus it is not surprising that the prediction scores which come from a model that depends on time are particularly helpful in differentiating the two DGMs.  In other words, the prediction scores reveal that the behavior of participants when there are 0\% bots as compared to 50\% bots differs most strongly with respect to the number of rounds; there is not a strong difference in regards to the participants' previous decisions or the average previous decision. 

\begin{figure}
\scriptsize $\hspace{4mm}$ Raw Simulated Data $\hspace{55mm}$ Model 1\\
\includegraphics[width=.49\textwidth,page=2]{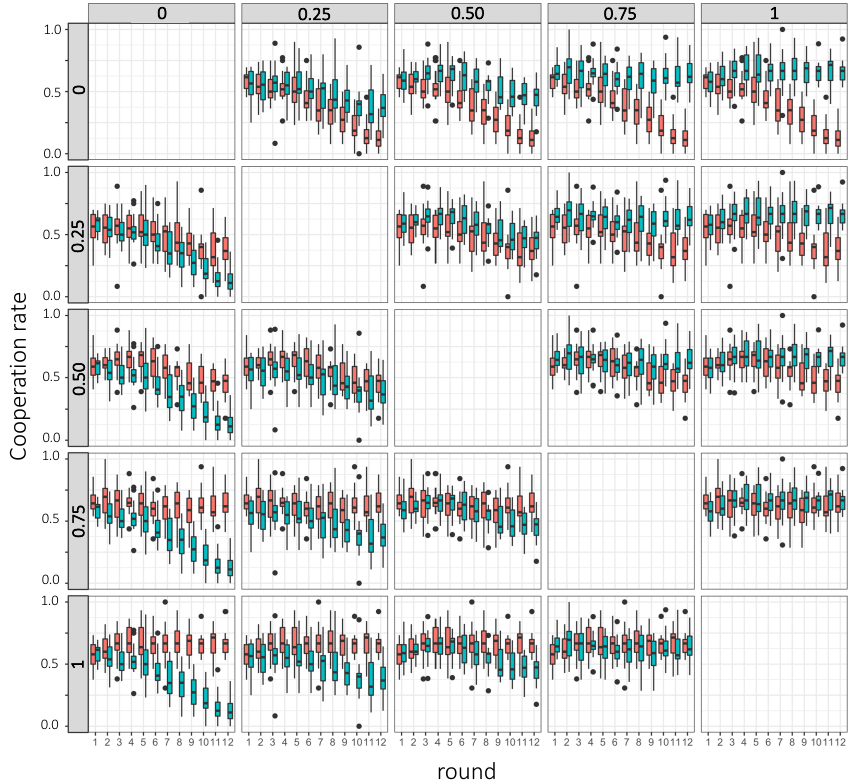}
\includegraphics[width=.49\textwidth,page=3]{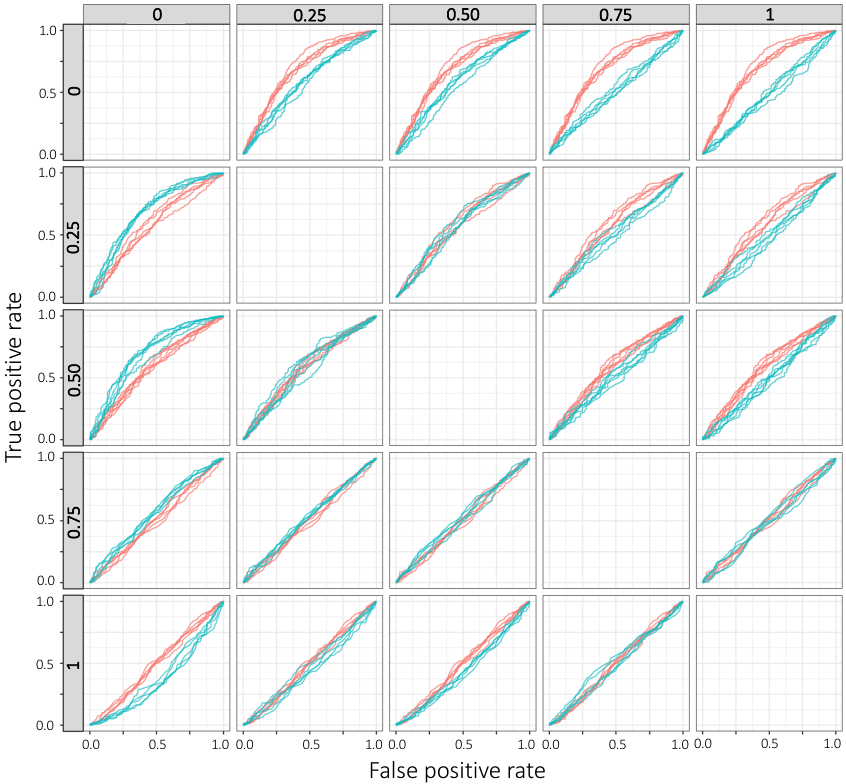}

\scriptsize $\hspace{4mm}$ Model 2 $\hspace{70mm}$ Model 3\\
\includegraphics[width=.49\textwidth,page=4]{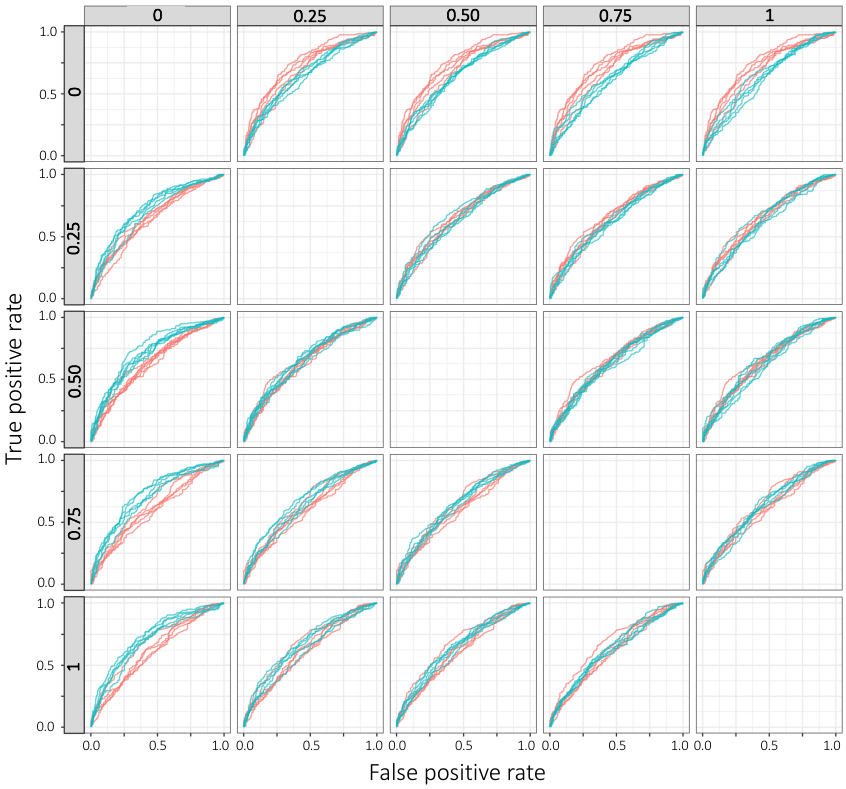}
\includegraphics[width=.49\textwidth,page=5]{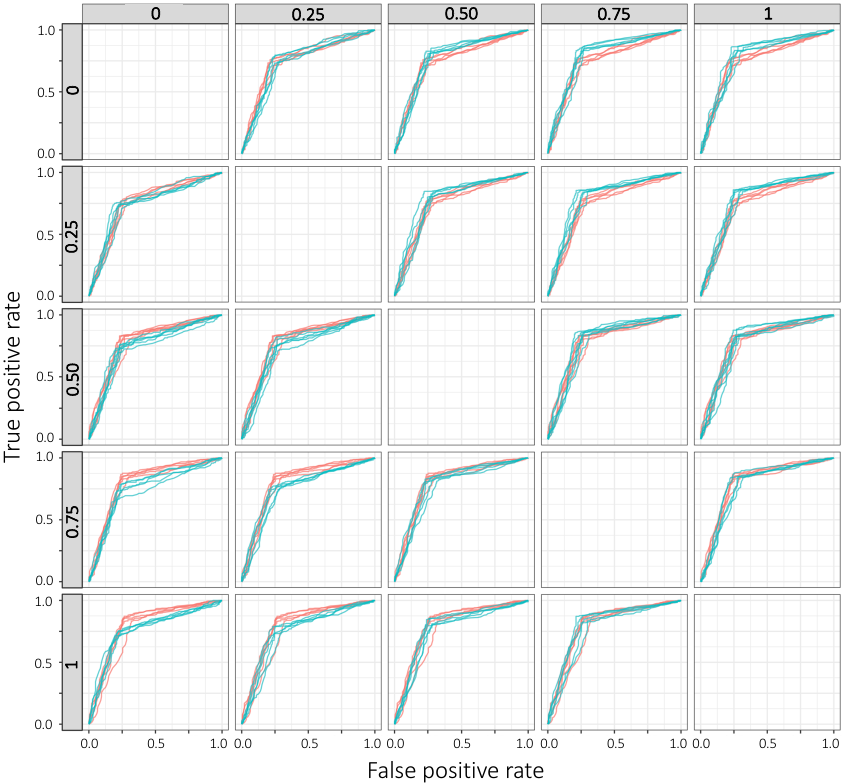}
\center{
\includegraphics[width=.4\textwidth,page=2]{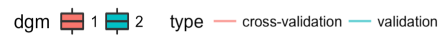}}
\caption{ROC curves from prediction scoring to compare across experimental conditions for $\pi \in \left\{0, 0.25, 0.50, 0.75, 1 \right\}$.  The panel on the top left contains boxplots of the cooperation rate across all cohorts and individuals by round, where the color represents the experimental settings.  Remaining panels display the prediction scoring ROC curves for subsets of the data from the cross validation (red) and validation (blue) routines, with each panel corresponding to a different researcher model.  Within each panel, the columns correspond to experimental conditions (values of $\pi$) for the first DGM and the rows correspond to the second DGM.}
\label{fig.rocAll}
\end{figure}

Finally, we examine the prediction scores across a range of experimental settings, making comparisons across $\pi = (0, 0.25, 0.50, 0.75, 1)$, in Figure \ref{fig.rocAll}.  Just as in Figure \ref{fig.rocExample}, we see that Model 1 is the most sensitive to differences across the experimental settings.  Further, as we might expect, as the distance between the experiments increases (in terms of $|\pi_1 - \pi_2|$), so too does the separation between the cross validation and validation ROC curves, especially for Model 1.  In other words, when the model is aligned with true differences between the data generating mechanisms, the distance between the cross validation and validation statistics reflects the true distance between the DGMs.

\paragraph{Summary}  This simulation study demonstrates that prediction scores go above and beyond traditional predictive accuracy measures, can be used to uncover important features of data generating mechanisms that differ across experimental settings, and can leverage competing predictive models to uncover different types of differences between the DGMs.  This is true even when the true data generating mechanism is unknown, as is the typical case in practice.  Here, the prediction scores discovered that the impact of round number or time in the game is best aligned with true differences between bot and human behavior.  This is reassuring since we can verify this effect by examining boxplots of the cooperation rate by round across each setting (e.g., the leftmost panel in Figure \ref{fig.rocExample}).  

\subsection{Estimating prediction score accuracy}
\begin{figure}
\centering
  \includegraphics[width=.9\linewidth,page=6]{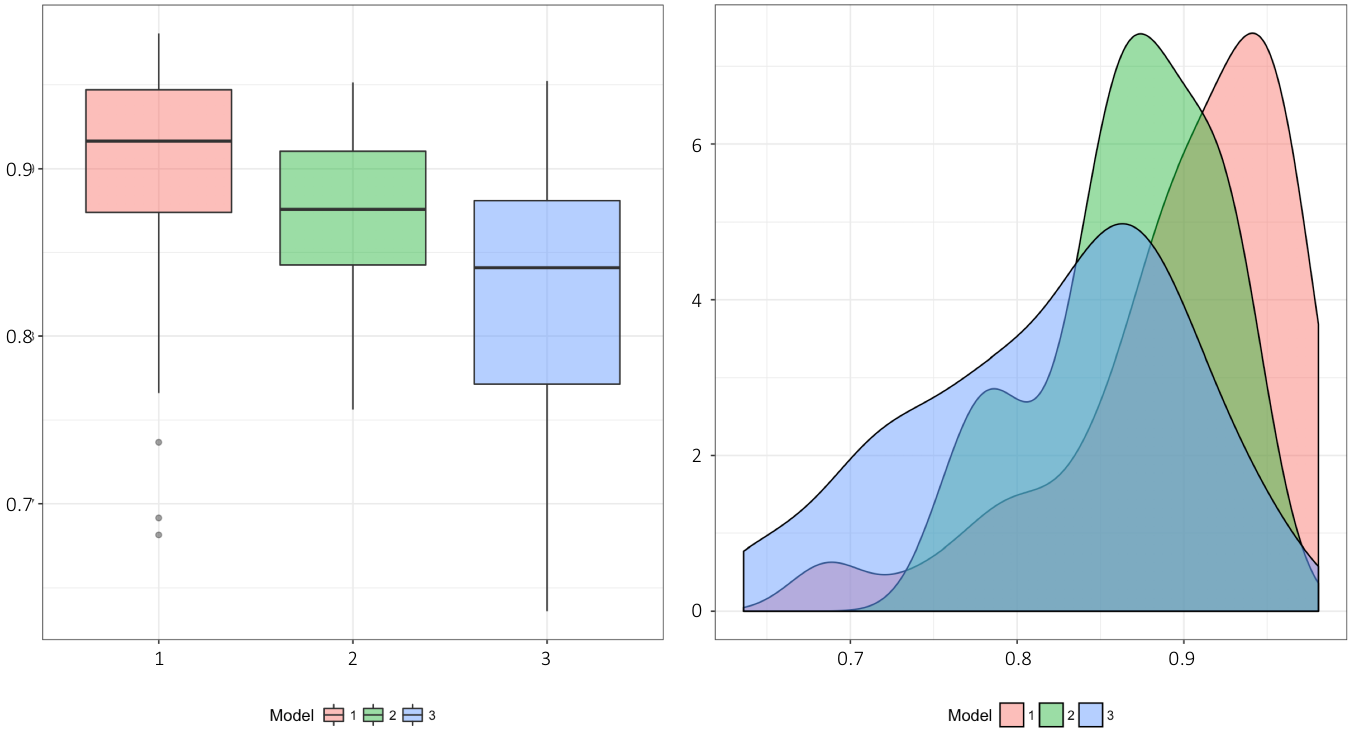}
  \caption{Distance correlations for prediction scores, based on 100 repetitions of simulated experimental data.  In each simulation, predictions are scored according to the predictive researcher models described earlier in Section \ref{sec.SimStudy} and by calculating Kolmogorov-Smirnov statistics that compare the empirical distributions of cross validation and validation AUC statistics.  The left panel displays these results as boxplots; the right panel displays the same results as empirical density plots.} 
  \label{fig:psBigSim}
\end{figure}

In order to get a sense of how these prediction scores behave asymptotically, we repeat the above simulation study and examine the relationship between the true distance between DGMs and our prediction scoring estimates of that distance.  This requires defining a true distance between the data generating mechanisms.  Here, we simply use the difference between the percentage of bots, $|\pi_i - \pi_j|$.  We compare this true distance to the prediction scoring estimates of distance, which we calculate as Kolmogorov-Smirnov statistics \citep{kolmogorov_1933} that compare the distributions of AUC statistics for the ROC curves across cross validation and validation.  To evaluate whether or not the prediction scoring estimates are well-aligned with this measure of the true underlying distance, we calculate distance covariances \citep{szekely_rizzo_bakirov_2007}.  A distance covariance is a measure of dependence between two paired vectors that is capable of detecting both linear and nonlinear associations.  If the vectors are independent, then the distance covariance is zero.  We can treat each repetition of the above simulation study (where we compute prediction scores across all possible pairs of $\pi$) as a sample which gives rise to a vector of prediction scoring distance estimates.  Then we examine the distribution of sampled distance covariances, as a function of the (researcher) model used to make predictions.  After repeating this simulation 100 times, we plot the distance covariances in Figure \ref{fig:psBigSim}.  As expected, we see that on average Model 1 out-performs Model 2 which out-performs Model 3, in terms of how correlated the prediction scoring estimates of distances between the DGMs are with the true distance, as measured by the difference in the percentage of bot participants.  This indicates that, on average, the prediction scores can successfully detect important differences between the DGMs.

\section{Preregistered hypotheses in human behavior experiments} \label{sec.AppNGS2}

Recall from Section \ref{sec.OurApproach} that one important motivating example for prediction scoring methodology is in the case of evaluating preregistered hypotheses.  In this section, we briefly review the idea behind preregistered analyses and traditional evaluation approaches.  We evaluate a preregistered hypothesis against realized experimental data from Cycle 1 of the NGS2 program (see the Appendix for more details) from the research team led by scientists at Gallup \citep{gallup_cycle1}, demonstrating how prediction scores provide important advantages over traditional NHST procedures and that prediction scores can identify important differences between pilot and experimental data.

\subsection{Preregistration}

In a preregistered design, researchers prepare a detailed plan for all data collection, coding, and statistical analysis, along with the hypotheses and corresponding predictions regarding the study's results.  This plan is made publicly available (``registered'') in some way before (``pre'') any data collection or analysis, so that the researchers are held accountable to their preregistered plan,\footnote{This sort of preregistration does not preclude further exploratory analyses; the point of preregistration is not to restrict analyses but rather to provide more structure to analyses that are already planned.  For example, after data collection, a researcher may notice a pattern or posit a new explanation that motivates additional analyses.  Such additional exploratory data analysis (beyond preregistered plans) are generally desirable as they can lead to new discoveries or hypotheses and even inspire additional confirmatory research.} and the ``garden of forking paths'' can be safely avoided \citep{gelman_loken_2014}.
Preregistration ensures that in such settings where a $p$-value is useful, it can be interpreted correctly.  Many journals, across many disciplines, now encourage preregistered studies, in the form of registered reports (e.g., the neuroscience journal, \textit{Cortex}), and any study's preregistration materials can easily be made publicly available on sites like the Open Science Framework.\footnote{The preregistration materials corresponding to the study data used in the human behavior example discussed in Section \ref{sec.AppNGS2} are hosted on this site.}

As best we can tell, current practice for prediction scoring generally consists of making predictions in the form of directional hypotheses (in some cases, predictions for the relative effect size are also included) for parameters in a model that captures our beliefs about the true underlying DGM.\footnote{Other approaches include using Bayes factors or the small-telescopes approach \citep{simonsohn_2015}, though these methods seem far less popular than traditional NHSTs.}  These predictions are then typically assessed by fitting the model to the observed experimental data, performing the corresponding hypothesis test and checking for a significant effect.  

\subsection{Example:  Experimental human behavior data}

\paragraph*{Experimental setting} In this study, the Gallup team was interested in understanding the role of social networks in the public goods game \citep{ledyard_1995}. In this version, participants' contributions are split only among neighbors in their (possibly evolving) social network.  Experimenters randomly assigned participants to one of four conditions which determined the dynamics of the social network in the game: (1) static or fixed links, (2) random link updating, where the entire network is regenerated at each round, (3) strategic link updating, where a randomly selected actor of a randomly selected pair may change the link status of that pair.  The strategic link updating condition was further split into two categories: (a) viscous, where 10\% of the subject pairs were selected and (b) fluid, where 30\% of the subject pairs are selected.  We will be primarily interested in the impact of the fluid version of the strategic link updating condition, from here on called ``rapidly updating networks.''  The Gallup team used a logistic regression model to examine individuals' decisions (cooperation or defection) under a variety of experimental conditions.  The Gallup team's experiments were inspired by the experiments performed by \citet{rand_arbesman_christakis_2011} and whose data can serve as a set of preregistration pilot data.

\begin{figure}
\centering
    \includegraphics[width=\linewidth,page=1]{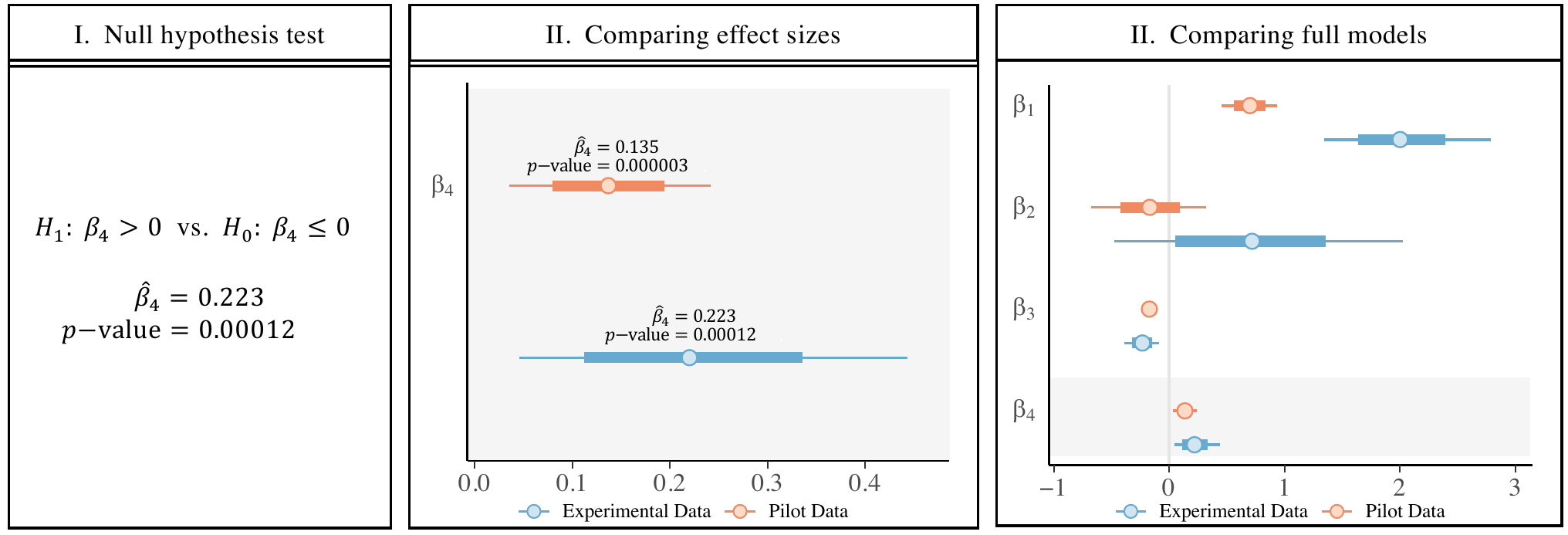}
  \caption{Various approaches for evaluating a preregistered hypothesis in the face of pilot and experimental data.}
  \label{fig:trad_analysis}
\end{figure}

\paragraph*{Traditional analysis} We consider the following hypothesis from the Gallup team's preregistration materials:  rapidly updating networks should support cooperation (across rounds of the game) more than any other condition \citep[see Hypothesis 1.4 of][]{nosek_spitzer_russell_etal_2018}.  The traditional approach to evaluating this hypothesis would be to specify a model which includes a parameter that compares the rapidly updating network condition to all other conditions and then to perform a null hypothesis statistical test.  Let $y_{it}$ represent the decision to cooperate ($y_{it}=1$) or defect ($y_{it}=0$) for participant $i$ in round $t$, which we can model as
\begin{align*}
    y_{it} &\overset{ind}{\sim} \text{Bernoulli} (p_{it}) \\
    \text{logit}^{-1}(p_{it}) &= \beta_1 + \beta_2 t + \beta_3 X_i + \beta_4 X_i t,
\end{align*}
where $X_i$ represents inclusion in the rapid updating network condition for participant $i$.  Then we are simply interested in testing:
\begin{align*}
    H_1 : \beta_4 > 0 \hspace{5mm} \text{ vs.} \hspace{5mm} H_0 : \beta_4 \le 0.
\end{align*}
In this case, the estimated effect size is about 0.22, with a $p$-value of 0.0001, and so one would traditionally conclude that the rapidly updating network conditions statistically significantly increase the likelihood of cooperation; see panel I of Figure \ref{fig:trad_analysis}.  For this analysis and all those to follow, we match the modeling strategy proposed in \citet{nosek_spitzer_russell_etal_2018} but will perform the analyses in a Bayesian setting.

However, this type of analysis does not incorporate the valuable information we have from the pilot data.  Perhaps we could compare the results of this hypothesis test to its analogue for the pilot data; see panel II of Figure \ref{fig:trad_analysis}.  In this case, the estimated effect size is about 0.14, with a $p$-value less than $10^{-5}$.  What does this allow us to say about how the pilot data compares to our experiment?  In our pilot data there is a significant increase in the likelihood of cooperative behaviors under the rapid updating network condition; we find the same effect in our experiment, with a larger effect size but with slightly less evidence that this network condition makes a significant impact.

But this analysis ignores other trends in the data that might differ across the two data sources.  Regression null hypothesis tests, like the ones above, are conditioned on all other terms (i.e., covariates) in the model.  And if we compare these other effects across the two data sources (see panel III of Figure \ref{fig:trad_analysis}), we see a difference in the estimates for the baseline rapid updating condition.  In our experiment, the baseline effect is positive and large and is significantly different than zero (though the $p$-value is on the larger side) whereas in the pilot data, the baseline effect is negative although not significant.  While participants seemed to be much more cooperative overall in our experimental data, the way the game progresses also seems to affect the decision to cooperate across these two settings.  From this analysis, it is unclear how these other differences between the experimental and pilot data might affect our hypothesis about the fluid network condition.  Other summaries of model fit are generally unhelpful here as well; take for example, AIC, which can compare non-nested models (like these) but the interpretation of comparisons across different responses (i.e., different observed outcome variables, $y$) is unclear.

\begin{figure}[t]
\centering

  \includegraphics[width=\linewidth,page=2]{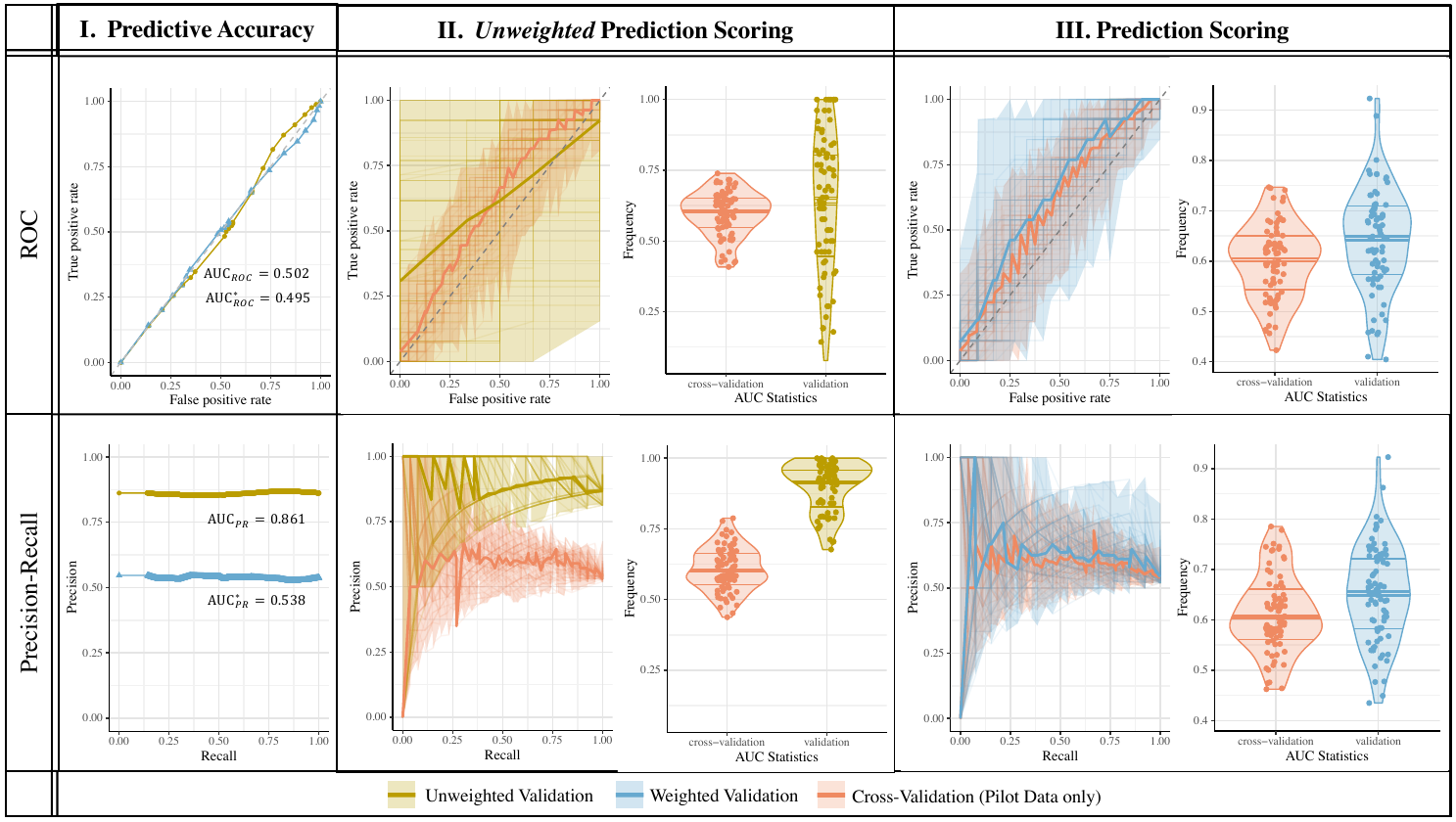}
  \caption{Prediction scores for Gallup's Cycle 1 Hypothesis 1.4 that rapidly updating networks support cooperation, relative to all other conditions.}
  \label{fig:ps_analysis}
\end{figure}

So, how does the pilot data differ from our experiment?  And how can we summarize these differences in a more holistic way that accounts for all trends related to our hypothesis about the rapid network condition?  As in Section \ref{sec.SimStudy}, we can consider traditional predictive accuracy metrics in the context of validation.  As in Section \ref{sec.SimStudy}, we will consider ROC and precision-recall curves,\footnote{The precision-recall curve is preferred over the ROC curve when the data are imbalanced, typically when there are many more $0$'s than $1$'s \citep[see][for more discussion]{davis_goadrich_2006}.  This is not the case here, since there are $n_1 = 3876$ observations in the pilot data and the average decision to cooperate is $p_1 = 0.53$.  Compare this to the experimental data, with $n_2 = 1192$ and $p_2 = 0.86$.} and the corresponding area under the curve statistics.  In this case, ROC seems to indicate that our model is little better than random guessing; see the yellow lines in panel I of Figure \ref{fig:ps_analysis}. At first the precision-recall curve appears to provide better news, but it is sensitive to differences in the proportions of $1$'s in the data.  We also noticed this difference between the pilot and experimental data in our comparison of the full logistic regression models in panel II of Figure \ref{fig:trad_analysis}.  So, to fairly assess the predictive accuracy of a model across two datasets, we need to ensure that the baseline rates are comparable.  To accomplish this, we resample the experimental data so that the baseline rates across the datasets are matched (equivalent results can be obtained via reweighting).  After this adjustment, the precision-recall curve seems to agree with the ROC in indicating that our model trained on the pilot data does not do a great job of predicting the experimental data; see the red lines in panel I of Figure \ref{fig:ps_analysis}.  Is that because our model does not represent the experimental data well (i.e., returning to Figure \ref{fig:flowAb}, the pilot and experimental data are clearly different and $\Delta_{DGM}$ is large)?  Or could it be that our model doesn't represent either the pilot data or the experimental data well (i.e., our model doesn't represent the DGM family well, and so can't tell us much about $\Delta_{DGM}$, the distance between the pilot and experimental data)?  To answer these questions, we need to be able to assess how well our observed pilot and experimental data represent the underlying DGMs; we need to represent the variability across datasets generated from the same DGM.  But these curves and any resulting analyses are conditioned on the particular observed (pilot) data.  We have no way of understanding the inherent variability in these types of summaries.  In fact, this is the case for all of the traditional analyses investigated thus far;  null hypothesis tests, effect sizes and predictive accuracy measures are all conditioned on the particular set(s) of observed (training and testing) data.

\paragraph*{Prediction scoring details} Recall from Figure \ref{fig:flowAb}, prediction scores require the following practical considerations: the predictive model, the loss statistic to compare predictions to realized data, and the subsampling method for the cross validation and validation routines.  As mentioned above, we fit Bayesian logistic regression models from the Gallup team's preregistration materials and we calculate ROC and precision-recall curves along with their corresponding AUC statistics; this means that we are considering two different ways of scoring the predictions (in practice, there may be many appropriate loss statistics).  Data subsets are created as random subsamples containing roughly $50$ and $25$ observations in training and testing sets, respectively.  As discussed above, experimental data are resampled so that baseline rates across all datasets are comparable, which ensures that precision-recall curves can be accurately compared.

\begin{figure}[t]
\centering
  \includegraphics[width=\linewidth]{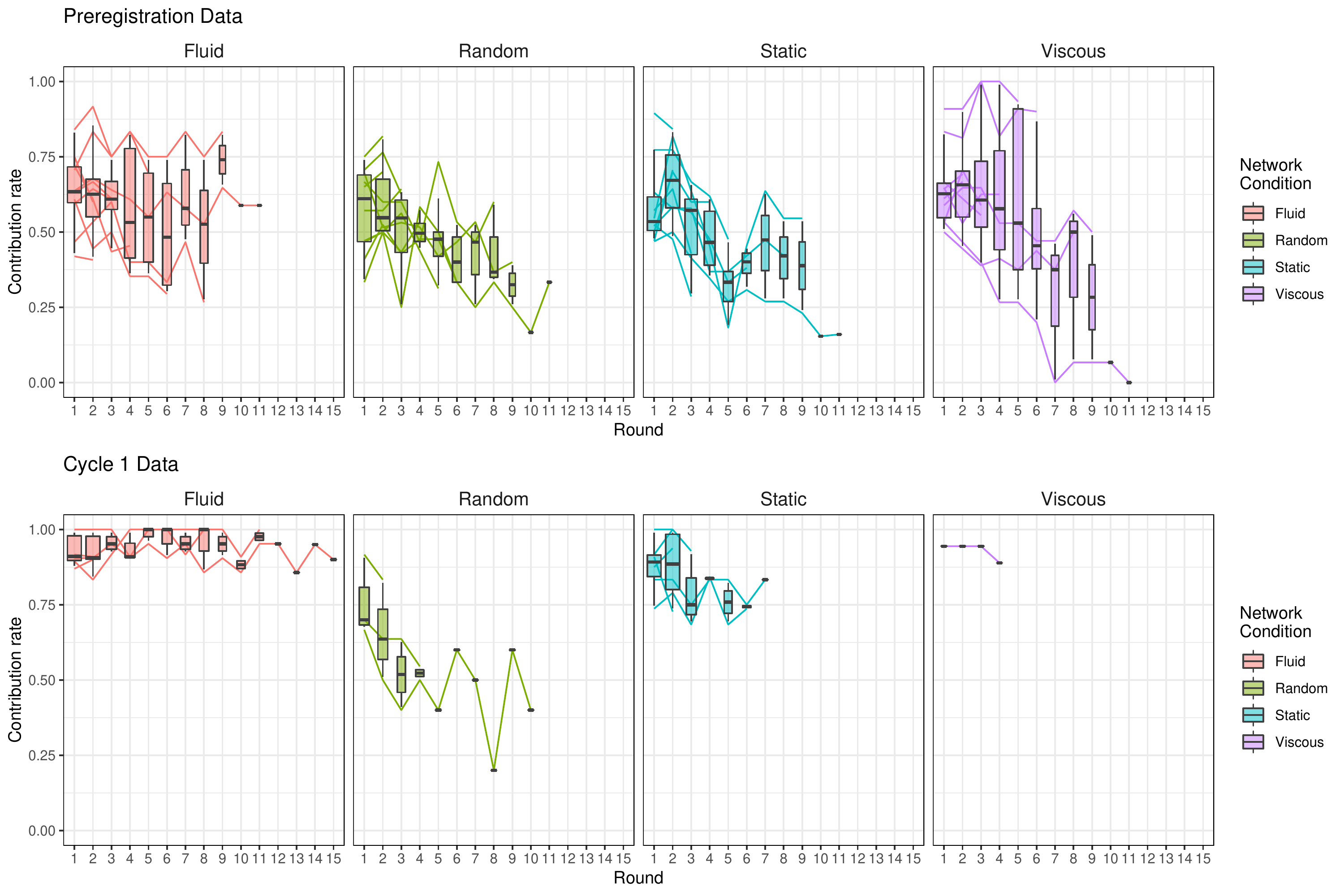}
  \caption{Average cooperation levels across rounds of Gallup's Cycle 1 games.}
  \label{fig:Cycle1b}
\end{figure}

\paragraph*{Results} We are evaluating the hypothesis that rapidly updating networks support cooperation more than any other condition.  As in Section \ref{sec.SimStudy}, we begin with visual comparisons of the ROC and precision-recall curves in panel III of Figure \ref{fig:ps_analysis}.  In general, we see little separation between the cross validation and validation curves, indicating that the underlying DGMs are likely to be similar; this matches the conclusion from our NHSTs discussed above.

However, the prediction scores additionally reveal that the researchers' logistic regression model surprisingly is a slightly better fit to the experimental data (particularly in terms of ROC).  This is somewhat surprising, since in most cases we expect a model to do a good job of predicting the data it was trained with, which would result in cross validation curves that look better then validation curves.  This indicates that there is less variability in individuals' behavior in the experimental data than in the preregistration data.  In a sense, subjects in the Gallup experiment are acting in more predictable ways than the subjects in the pilot data.  This conclusion is reached through the lens of the predictive model---here, logistic regression---and indicates that the experimental data are a better fit for this modeling framework (as opposed to the pilot data).  That is, the prediction scores identify a way in which the underlying DGMs differ; using the notation in Figure \ref{fig:flowAb}, the underlying DGM for the experimental data, $F'$, is more similar to a logistic regression model, $\hat{f}$, than the underlying DGM for the pilot data, $F$.  And in fact, if we simply examine summary statistics of the in-game decisions themselves, we can see the same type of pattern.  In Figure \ref{fig:Cycle1b}, we provide boxplots of individuals' average cooperation levels across rounds of the game, where each color corresponds to a different link-updating experimental condition.  Comparing the preregistration data (top row) to the experimental data (bottom row), we see that the boxplots are drastically narrower, especially in the fluid network condition, indicating that there is less variability in participant behavior.  Importantly, this is not a difference between the two datasets that would have been picked up by the traditional prediction score, the $p$-value for the NHST associated with $\beta_4$.

The validation curves themselves vary more, across training sets.  This is even more obvious if we compare the loss statistics, the AUC statistics for each of the curves in these figures.  Again, we begin with a visual comparison of the distribution of AUC statistics from cross validation to those from validation; see panel III of Figure \ref{fig:ps_analysis}.  Some of this could be due to the difference in sample sizes, particularly the difference in the sizes of the testing sets.  In this analysis, we've chosen $N/K$, the size of the training sets, to be 50 observations;  thus, with $N = 3876$ in the pilot data and $N' = 1915$ for the resampled experimental data (we need to resample here so that the baseline rate is comparable across both datasets), this yields testing sets of roughly 50 and 25 observations for the cross validation and validation routines, respectively. The preferred loss statistic will often dictate a minimum size for these testing sets.  So, here, it is possible that some of this increased variability in the validation curves is due to the smaller testing set size.  To counter this effect, we could consider resampling the experimental data further, to match the size of the pilot data.  But this runs the risk of going too far in the opposite direction, as resampled data tends to underestimate sampling variability.


We have created similar summary measures for the unweighted version of the prediction scores; see panel II of Figure \ref{fig:ps_analysis}.  As mentioned earlier, the precision-recall curves are much more sensitive to the difference in the baseline rates across the two datasets. This is not an obvious consequence of the precision-recall curves themselves, so care must be taken in fine-tuning the prediction scoring algorithm (i.e., choosing an appropriate loss statistic and possibly adjusting the resampling method) in order to obtain prediction scores that capture interesting features of the underlying DGMs, rather than more basic differences of the observed data (e.g., baseline cooperation rates).  In this sense, it can be helpful to consider competing loss statistics, as shown here with AUC statistics for both ROC and precision-recall curves.

Overall this application serves as an illustration of how our prediction scoring method can be used to evaluate preregistered hypotheses and to enable interesting scientific insights.  Here, while the NHST confirmed our preregistered hypothesis (the fluid network condition does support cooperation), prediction scoring nicely complemented this analysis by confirming that the DGMs appear to be similar and additionally highlighting differences in the underlying DGMs through the lens of our logistic regression model:   participants in the experimental data acted more predictably than participants in the pilot data, particularly in the fluid network condition.  In other words, the experimental data exhibited less variation in participants' decisions to cooperate over the course of the game (after accounting for differences across rounds).

\section{Discussion} \label{sec.Disc}

A natural version of the {\em prediction scoring} question might be phrased as follows: are these experiments or realizations products of the same DGM or are they distinct in some way?  While this is certainly a natural question, it is ill-posed for most experimental research settings.  In almost all cases, the DGMs do in fact vary across experiments or settings (e.g., from preregistration to observed data), even if only slightly.  Instead, we have focused on answering the following question: How much do the data generating mechanisms differ across settings?  To this end, our methodology provides prediction scores on a continuous scale; these scores can be viewed as estimates of the distance between our prior beliefs and reality.  Thus, they provide a quantitative measure of how well preregistered predictions align with reality rather than relying on the simple binary detection of a significant effect.

This methodology that utilizes cross validation and model-based predictions to solve a common problem in applied statistics research:  the evaluation of differences between DGMs. In practice, the DGMs may represent related experiments, different settings or conditions within a single experiment, or preregistered hypotheses and realized observational data.  We argued that comparing DGMs should move away from the simplistic binary question of whether or not the DGMs are equal and instead our prediction scores enable a quantifiable measurement of differences between DGMs.  In an application to human behavior experiments, we demonstrated how the prediction scores can be used to evaluate preregistered hypotheses and for a set of simulated experimental data, we demonstrated how these scores can detect important differences between experimental settings.  We also provided some intuition for the probabilistic behavior of these scores in an asymptotic regime.

Our application to the NGS2 project highlights the important role that our proposed prediction scores can play in light of the replication crisis.  The majority of statistical concerns arising from the replication crisis fall into one of the three following topics: (1) selection bias (i.e., only research with small $p$-values is submitted and published), (2) insufficient power (i.e., $p$-values that correspond to small samples are not to be trusted), and (3) how statistical inference differs from scientific inference \citep{colling_szucs_2018}.  Our prediction scoring methodology attempts to address this third issue but cannot fix issues arising from selection bias or insufficient power; these concerns still need to be carefully monitored.  Much of the discussion around this issue has focused on reconciling the questions scientists want answers to with the questions that traditional $p$-values are equipped to provide. The proposed prediction scores are not $p$-value replacements; they are designed to answer yet another type of important scientific question:  how do we quantify differences between DGMs?  Or in the preregistration setting, how do we compare preregistered hypotheses to realized experimental data in a meaningful way?  At its core, prediction scoring is a method for quantifying the distance between competing DGMs, one representing our prior/preregistered beliefs and one representing the realized experimental data.

In the examples in this paper, we focused on (visually) identifying differences between the distributions of loss statistics.  In practice, visual inspection of plots always has a subjective component; these plots also represent empirical samples of the full loss statistic distribution (across infinitely many subsampled testing/training sets).  Even when the prediction scoring procedure itself is preregisetered, this interpretation – which cannot be fully pre-registered because it depends on realized experimental data – involves some amount of subnjectivity or researcher degrees of freedom.  However, as mentioned above, our proposed approach is not designed as a $p$-value replacement, but rather as an important tool to take full advantage of the unique statistical problem presented by the preregistration setting; to provide researchers with a method to make these valuable comparisons between pre-experimental data and realized observed data in a way that goes beyond simple NHSTs.

As proposed, our prediction scores are not proper distances.  In this sense, we don’t expect them to be symmetric.  It matters which dataset plays the role of $\tau$ (i.e., is used to create a baseline, via cross validation) and which plays the role of $\tau'$ (i.e., is used in validation).  In the motivating example of the preregistration setting and as discussed in the real data example in Section \ref{sec.AppNGS2}, there is a natural directionality.  However, in other settings this may not be the case.  For more general cases, perhaps a symmetrized distance could be created by swapping the roles of $\tau$ and $\tau'$, and averaging the resulting sets of scores in some fashion.

AUC statistics were utilized in all of the case studies investigated here.  It is worth noticing that  the  traditional  AUC  statistic  is  not  defined  for datasets consisting  of only one observation. This means that leave-one-out cross validation is not feasible for this choice of loss statistic. In our applications, we accommodated for this by choosing $k < n$ for the $k$-fold cross validation routines in  our  prediction  scoring  method; this ensures  that  there  are  sufficient  data  points  in  the  holdout  sample.  Alternatively, other approaches for calculating the ROC or precision-recall curves in small sample settings could be incorporated \citep[e.g.,][]{yousef_wagner_loew_2005}.

Both the application and simulated experiments involve rather simple DGMs and predictive models (i.e., both involve logistic regression models).  This is by design, as we would like to be able to verify that the prediction scores are capable of detecting meaningful differences between the DGMs.  However, we recognize that many applied problems hope to address more complicated DGMs and require more complex modeling strategies.  This is an important avenue for future research; as discussed earlier, prediction scores are well-equipped to evaluate complex DGMs since they do not rely on the choice of a single parameter or summary statistic upon which to base the evaluation of differences between the DGMs.  For example, in planned future experiments from the NGS2 program, researchers have preregistered hypotheses based on Gaussian process models for the DGMs under study.  We hope to utilize prediction scoring in these settings, both to evaluate the preregistered hypotheses and to uncover interesting differences between experimental settings.  For example, in some experiments, bots (computer agents whose behavior is algorithmically determined) will participate alongside human participants;  we hope to use prediction scores to detect scientifically interesting differences in the observed (highly nuanced) human behavior patterns between human-only and human-and-bot experimental conditions.

\section*{Funding}
This material is based on and supported through research sponsored by the Defense Advanced Research Projects Agency (DARPA) agreement number D17AC00001. The content of the information does not necessarily reflect the position or the policy of the Government, and no official endorsement should be inferred.

The authors have no relevant financial or non-financial interests to disclose.

\section*{Data and Code}
Experimental data analyzed in Section \ref{sec.AppNGS2} is available on the Open Science Framework through associated GitHub links \citep{gallup_cycle1}. The simulated data and all code used in this paper is available
in an additional public Github repository \citep{predscore_github}.

\appendix

\section{The NGS2 Program}
The NGS2 program is a multi-phase methodologically-focused effort to develop a fundamental reimagining of the social science research cycle \citep{nosek_spitzer_russell_etal_2018}.  In each phase, distinct research teams conducted unique experimental social science studies regarding a shared research question. Prior to any data collection, each team was required to document all preregistration materials, including predictions for study outcomes \citep[for more detailed descriptions of each team's planned and completed research, see the preregistration materials which have been made publicly available on the Open Science Framework;][]{nosek_spitzer_russell_etal_2018}.  The program also required that each team's preregistered hypotheses and resulting final analyses be evaluated by external non-team members, which included the authors of this paper.  It is precisely in this context that our proposed prediction scoring methodology was developed.

In the first cycle of the program, research teams focused on explaining and predicting the emergence of collective identities. Collective identity refers to the way individuals perceive themselves in their environment with respect to the various groups they may belong to and how they subsequently take collective action or display collective behaviors related to this identity.  In Section \ref{sec.AppNGS2}, we examine how well the preregistered hypotheses align with the realized experimental data from the research team led by scientists at Gallup \citep{gallup_cycle1}.  The Gallup team's experiments were designed to mimic those of \citet{rand_arbesman_christakis_2011}; these experiments serve as pilot data and helped to formulate the team's preregistered hypotheses.

In the second cycle of the program, research teams designed experiments and analyses to study the emergence of group innovation in the face of competition.  The design of the simulation study described in this section is inspired by the proposed Cycle 2 experiments of the research team led by scientists at the University of Pennsylvania \citep{penn_cycle2}.  These experiments examine human behavior in the face of computer generated participants.

\section{Additional background}
\label{subsec.LitRev}

In this section, we provide relevant background information on the existing statistical methods which motivate our proposed prediction scoring framework, which will be fully developed in Section \ref{sec.cvPredScore}.  As outlined in the previous section, our approach is inspired by existing tools for measuring predictive accuracy (Section \ref{subsec.PredAcc}) and proposes a framework based on cross validation (Section \ref{subsec.CV}).  We also offer a discussion of how our proposed approach relates to recent work in algorithm validation (Section \ref{subsec.Algs}), which motivates strategies adopted in our proposed framework. 


\subsection{Scoring rules}
\label{subsec.PredAcc}
Scoring rules measure the agreement between a probabilistic forecast (a predictive probability distribution over future quantities or events of interest, such as a posterior predictive density from a Bayesian analysis) and an observation \citep[][provides a nice summary of recent research in this area]{gneiting_katzfuss_2014}.  This literature provides a sound framework for comparing probabilistic forecasts or predictions (such as from preregistration materials) to observed data, where each competing forecast could correspond to different modeling choices or assumptions.  The diagnostic tools and recommendations for scoring rules---e.g., checking for uniformity in histograms (or empirical CDFs, if the sample size is small) of the PIT (probability integral transform) values \citep[this idea can be traced as far back as][and perhaps earlier]{rosenblatt_1952,pearson_1933}---are predictive and thus enable the comparison of non-nested, highly diverse models fit to common data.  For example, \citet{pers_albrechtsen_holst_etal_2009} use strictly proper scoring rules to select between competing machine learning models.  However, since these tools were developed from the perspective of forecast selection (e.g., choosing the best forecast from among a group of competing forecasts), each set of resulting diagnostic measures is necessarily model-based in that any diagnostic plot or set of scoring rules depends on the model assumptions used to create the probabilistic forecast.  This complicates the interpretation of the scores or diagnostics in regards to true underlying differences between the DGMs, since they can detect differences between the DGMs but are also designed to measure differences between the model and the DGM, which may be attributable to model fit issues.  As mentioned earlier, our proposed prediction scoring approach will use cross validation to help normalize for model fit issues, but many of the proposed scoring rules could be incorporated in our proposed method.

\subsection{Cross validation}
\label{subsec.CV}

Cross validation, particularly for Bayesian analyses, has been an active research area in recent years.  Summary statistics for comparing Bayesian models can be motivated by estimation of out-of-sample predictive accuracy \citep[see][for a thorough review, from a formal decision theoretic perspective]{vehtari_ojanen_2012}, which  is one of the goals of cross validation as well. \citet{gelman_hwang_vehtari_2014} provide a review of some model comparison summary measures, including AIC, DIC, WAIC, in the context of Bayesian model comparison.  As opposed to exact leave-one-out cross validation (LOOCV), each of the Bayesian model summary statistics utilize the full predictive density and perform an adjustment (e.g., importance sampling, or division by an appropriate variance) to remove the effect of over-fitting, since no data was actually held out.  The authors conclude the paper by citing cross validation as their preferred method for model comparison, despite its high computational cost and requirement that data can be easily partitioned (i.e., partitioning is often not straight forward for dependent or hierarchical data).  In this line of thought, \citet{vehtari_gelman_gabry_2017} develop an approximate version of leave-one-out cross validation which implements Pareto smoothing of the importance sampling weights to improve robustness to weak priors or influential observations.  \citet{li_qiu_zhang_etal_2016} develop a version of cross validation that can be applied to models with latent variables, which relies on an integrated predictive density.  In applications with competing probabilistic forecasts, \citet{held_schrodle_rue_2010} compare software fitting algorithms using approximate cross validation and many of the diagnostic plots mentioned by \citet{gneiting_balabdaoui_raftery_2007}.  Finally, \citet{wang_gelman_2014} and \citet{millar_2018} address the problem of appropriate data partitioning and out-of-sample prediction error estimation for multilevel or hierarchical model selection using cross validation and predictive accuracy.  \citet{wang_gelman_2014} highlight the fact that model selection can be largely based on the size and structure of the hierarchical data.

This line of research, and its proposed improvements and extensions of cross validation in various Bayesian settings, can certainly be incorporated in the prediction scoring methodology that we propose.  Our contribution will be to expand this literature, from the perspective of the preregistration setting; we formalize the use of cross validation to appropriately adjust agreement measures between preregistered predictions and realized observations.  In other words, we will recommend a unique combination of cross validation \textit{and} external validation to provide meaningful prediction scores.

\subsection{Algorithm checking}
\label{subsec.Algs}

Although perhaps not obvious at first glance, recent proposals for checking algorithms of Bayesian model fitting software \citep{cook_gelman_rubin_2006, talts_betancourt_simpson_etal_2018} can provide insights in the prediction scoring setting.  These proposals recommend simulating fake data conditional on random draws from the prior distribution, running the model fitting software to obtain draws from the posterior distribution, and using a summary measure to diagnose the alignment between posterior samples and the random draws from the prior distribution.  Based on the self-consistency property of the marginal posterior and the prior distribution, these draws should be indistinguishable from one another.  To diagnose this alignment, \citet{talts_betancourt_simpson_etal_2018} suggest computing rank statistics, comparing the random draw form the prior distribution to the posterior distribution based on that particular draw.  The authors suggest looking at histograms of these quantiles, demonstrating that if the software is working correctly, the quantiles should follow a discrete uniform distribution.  
In the prediction scoring setting, we can think of this software-checking methodology as a special case where the chosen modeling strategy matches the underlying DGM exactly (i.e., there are no model fit issues whatsoever).  We will borrow ideas from this methodology, such as the use of empirical quantiles and rank statistics to compare DGMs (or distributions) through samples drawn from them.

\def\url#1{}
\bibliography{compare}

\end{document}